\documentclass[prb,aps,twocolumn,english,floatfix,superscriptaddress]{revtex4-2}

\usepackage{epsfig,graphicx,psfrag,amsmath,amssymb,float}

\usepackage{mathtools}

\usepackage{amsmath}

\usepackage{amssymb}

\usepackage[dvipsnames]{xcolor}

\usepackage[
	colorlinks=true,
	linkcolor=blue,
	anchorcolor=blue,
	citecolor=OliveGreen,
	urlcolor=NavyBlue
]{hyperref}

	\newcommand{\bra}[1]{\langle{#1}|}

	\newcommand{\ket}[1]{|{#1}\rangle}
	
\begin{document}


\title{Fractons from a liquid of singlet pairs}

\author{Hernan B. Xavier}
\affiliation{Departamento de F\'isica Te\'orica e Experimental, Universidade Federal do Rio Grande do Norte, 59072-970 Natal-RN, Brazil}

\author{Rodrigo G. Pereira}
\affiliation{Departamento de F\'isica Te\'orica e Experimental, Universidade Federal do Rio Grande do Norte, 59072-970 Natal-RN, Brazil}
\affiliation{International Institute of Physics, Universidade Federal do Rio Grande do Norte, Natal, Rio Grande do Norte 59078-970, Brazil}

\date{\today}

\begin{abstract}

Fracton phases of matter feature a  variety of exciting phenomena stemming from  the restricted mobility of their quasiparticles. Here we consider a model of interacting electrons in one dimension that describes hopping of spin-singlet pairs and obeys both charge and dipole conservation laws. The model contains Bethe ansatz integrable sectors which allow us to solve the ground state and calculate  the exact  spin and single-electron excitation gaps. We observe hallmarks of fractonic behavior, including localization of single-electron excitations and propensity to clustering. Our results demonstrate the important role of the dipole moment conservation law in a simple model of spin-1/2 fermions.

\end{abstract}

\maketitle


\section{Introduction}

Fractons are quasiparticles which cannot move in isolation, but are allowed to move by forming certain bound states \cite{Chamon2005,Vijay2016,Pretko2017a,Prem2018,Sous2020PRB,Kumar2019,Sous2020}. Fractonic behavior may arise from higher-moment conserved charges, such as the dipole moment in tensor gauge theories \cite{Pretko2017a}. While the glassy dynamics of fractons \cite{Chamon2005} has drawn interest as a potential platform for robust quantum information storage \cite{Haah2011,Bravyi2013,Terhal2015}, their fundamental properties have also led to deep connections with a wide variety of concepts, ranging from many-body localization \cite{Prem2017,Sala2020,Khemani2020} to gravity and holography \cite{Pretko2017b,Yan2019}. For a  broad perspective on fracton phases of matter, we refer the reader to the reviews in Refs. \cite{Nandkishore2019,Pretko2020}.

While the original studies focused on spin models in three dimensions \cite{Chamon2005,Haah2011,Vijay2016}, the importance of searching for fracton phenomenology in one dimension has recently been underscored   \cite{Sous2020PRB,Pai2020}. The restriction to one spatial dimension opens the possibility of employing exact analytical methods, such as Bethe ansatz solutions for integrable models \cite{Bethe1931,Sutherland2004}. Noteworthy examples of Bethe ansatz solvable models include the Hubbard chain \cite{Lieb1968,Essler2005}, the supersymmetric $t$-$J$ model \cite{Schlottmann1987,Essler1992}, and some correlated hopping models \cite{Bariev1994,Alcaraz1999}. Beyond integrability, in one dimension one can also resort to powerful numerical techniques to study static as well as dynamic properties \cite{Schollwoeck2011}. In fact, the nonequilibrium dynamics governed by kinetically constrained hopping  of particles in one dimension has received a great deal of attention in the context of weak ergodicity breaking from quantum many-body scars \cite{Sala2020,Khemani2020,Hudomal2020,Moudgalya2020,Serbyn2020}.

In this paper we examine an interacting  one-dimensional model that describes a liquid of singlet pairs in which single electrons behave as fractons. This singlet pair liquid bears a resemblance to doped quantum dimer models which may be relevant to the theory of high-$T_c$ superconductors in higher dimensions \cite{Rokhsar1988,Poilblanc2008,Punk2015}. As a matter of fact, the origin of the model   traces back to the strong coupling limit of the Hubbard model, as the pair hopping Hamiltonian  in Eq. (\ref{eq:model}) below corresponds to the three-site term generated at the same order in perturbation theory as the Heisenberg exchange interaction \cite{Chao1978,Huang1987,Fazekas1999}. The complete model  including   electron hopping with amplitude $t$, the  nearest-neighbor exchange interaction $J$, and the pair hopping term with coupling constant  $\alpha J$ has been dubbed $t$-$J$-$\alpha$ model \cite{Ammon1995,Coulthard2018}.  The latter  has been analyzed using mean-field approximations \cite{Huang1987}, exact diagonalization on small chains \cite{Ammon1995,Saiga2002}, and, more recently, density matrix renormalization group techniques \cite{Coulthard2017,Coulthard2018,Gao2020}. Remarkably, Floquet engineering can be used to  enhance the pair hopping term \cite{Coulthard2018,Gao2020}.  To the best of our knowledge, the pure singlet-pair hopping model in one dimension was first studied by Batista \textit{et al.} \cite{Batista1995}, who, guided by numerical solutions, calculated  the exact ground state energy and found an energy spectrum characteristic of a Luther-Emery liquid, \textit{i.e.}, a Luttinger liquid with a spin gap but no charge gap \cite{Luther1974}.

We revisit  the singlet-pair hopping model in the context of fracton phases of matter.  We extend previous work \cite{Batista1995} by providing a Bethe ansatz solution in  integrable sectors  that include  the ground state and spin-triplet excitations.  In the sector in which all electrons are bound into mobile singlets with charge $2e$, the gapless excitations are the bosonic charge modes of the singlet-pair liquid.  Triplet excitations are completely immobile and  amount to impenetrable barriers for the singlet pairs. By contrast, the sectors which contain isolated  electrons are not integrable. Our key contribution stems from the observation that, besides the standard charge conservation law, the Hamiltonian also commutes with a dipole-type operator. While electrons can move assisted by the hopping of an adjacent singlet pair, the conservation of the dipole moment implies that  single-electron wave functions are localized even in the absence of quenched disorder. Furthermore, we observe  a tendency towards  clustering of electrons due to an effective attractive interaction mediated by singlet pairs. We point out that perturbations to the pure singlet-pair hopping model, such as a magnetic field or a nearest-neighbor repulsive interaction, can close the gap for spin or  single-electron excitations, thereby driving transitions to yet unexplored phases with a finite density of fractons.

The rest of the paper is organized as follows. 
In Sec. \ref{sec:model-charges}, we introduce the model, discuss relevant conserved charges, including the dipole operator, and provide a classification for the elementary excitations.
In Sec. \ref{sec:ground-state}, we find the exact ground state by Bethe ansatz techniques and derive the low-energy effective theory describing a Luttinger liquid of charge-$2e$ particles.
Spin and single-electron excitation gaps are calculated in Sec. \ref{sec:excitations}.
In Sec. \ref{sec:localization}, we continue to explore single-electron excitations, laying emphasis on their localized character.
In Sec. \ref{sec:perturbations}, we consider the effects of a magnetic field and a nearest-neighbor repulsion.
We offer a summary and concluding remarks in Sec. \ref{sec:conclusions}.
Finally, Appendix \ref{app:hopping} contains an analysis of the role of single-particle hopping in the two-electron problem.


\section{Model and conservation laws}\label{sec:model-charges}

We investigate a one-dimensional lattice model of interacting electrons that describes hopping of spin-singlet pairs. We also assume there is an infinite on-site repulsion that rules out doubly occupied sites. The Hamiltonian is  
\begin{equation}\label{eq:model}
H=- \sum_{j}\mathcal{P}\left(b_{j-1,j}^{\dagger}b^{\phantom\dagger}_{j,j+1}+\text{H.c.}\right)\mathcal{P}.
\end{equation}
Here $b_{ij} = \frac{1}{\sqrt{2}}(c_{i\downarrow}c_{j\uparrow}-c_{i\uparrow}c_{j\downarrow})$ is the operator that annihilates a pair of electrons at sites $i$ and $j$ in a spin-singlet state and $\mathcal{P}=\prod_j(1-n_{j\uparrow}n_{j\downarrow})$, with $n_{j\sigma}=c^\dagger_{j\sigma}c^{\phantom\dagger}_{j\sigma}$ for $\sigma=\uparrow,\downarrow$, is the projection operator that implements the exclusion of double occupancies at every site. Despite the simple-looking structure of the Hamiltonian, we must be careful since the pair operators $b_{ij}$ do not obey canonical commutation relations. Rather, when they overlap, we have the commutator
\begin{equation}
[b^{\phantom\dagger}_{ij},b_{jl}^{\dagger}]=\delta_{il}\left(1-\frac{n_{j}}{2}\right)-\frac{1}{2}\sum_{\sigma}c_{l\sigma}^{\dagger}c^{\phantom\dagger}_{i\sigma},
\end{equation}
where $n_{j}=n_{j\uparrow}+n_{j\downarrow}$.  In terms of  electron operators, 
the Hamiltonian reads
\begin{align}\label{eq:model-c-op}
H = -\frac{1}{2}&\sum_{j,\sigma}\mathcal{P}\Big(c_{j-1,\sigma}^{\dagger}c_{j,\bar{\sigma}}^{\dagger}c^{\phantom\dagger}_{j,\bar{\sigma}}c^{\phantom\dagger}_{j+1,\sigma} \nonumber\\
&\qquad -c_{j-1,\sigma}^{\dagger}c_{j,\bar{\sigma}}^{\dagger}c^{\phantom\dagger}_{j,\sigma}c^{\phantom\dagger}_{j+1,\bar{\sigma}}+\text{H.c.}\Big)\mathcal{P},
\end{align}
where $\bar \sigma=\downarrow,\uparrow$ for $\sigma=\uparrow,\downarrow$, respectively. This is precisely the three-site term generated by perturbation theory in the strong coupling limit  of the Hubbard model with hopping parameter $t$ and on-site repulsion $U\gg t$ \cite{Chao1978,Huang1987,Fazekas1999,Ammon1995,Coulthard2018}.  In that case, the operator in Eq. (\ref{eq:model}) is obtained with coupling constant $\alpha J$ with $J=4t^2/U$ and $\alpha=1/2$. However, here we consider the limit of the pure pair-hopping model and set the coupling constant to unity.

\begin{figure}
\includegraphics[scale=0.5]{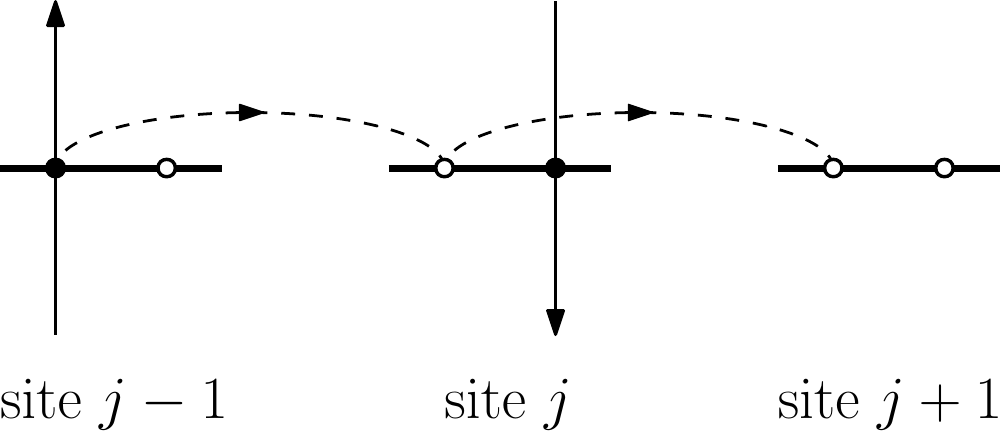}\vspace{.5cm}
\includegraphics[scale=0.5]{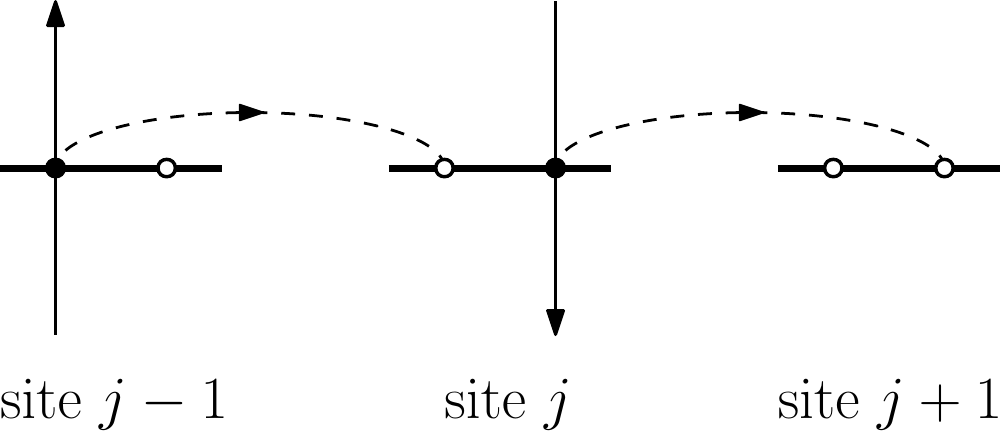}
\caption{Correlated hopping processes that enable the motion of a spin-singlet pair. The two processes amount to second-neighbor-hopping, with or without spin flip, conditioned to the presence of an electron with opposite spin in the intermediate site. }
\label{fig:singlet-hopping-mechanism}
\end{figure}

The three-site correlated hopping   processes are illustrated in Fig. \ref{fig:singlet-hopping-mechanism}. In the first process, associated with the first term in Eq. (\ref{eq:model-c-op}), an electron can hop to a next-nearest-neighbor site if the intermediate site is occupied by another electron with the opposite spin. In the second process, the second-neighbor hopping   is accompanied  by a spin flip for both electrons. Alternatively, we can think that the first electron hops to the intermediate site while the second electron hops to the third site. The interference between these two processes allows for  motion   only   if the electrons are in a singlet state, as made explicit in Eq. (\ref{eq:model}). Thus, this Hamiltonian enables a singlet pair to move, provided there is an adjacent empty site.

Let us now see what sort of conservation laws we may have. First, it is clear that both the total number of electrons $N=\sum_j n_j$ and total spin $z$-projection $S^z=\frac{1}{2}\sum_j(n_{j\uparrow}-n_{j\downarrow})$ commute with the Hamiltonian, and are thus good quantum numbers. 
We shall use them to label eigenstates of the Hamiltonian according to the eigenvalue equation
\begin{equation}\label{eq:schrodinger-eq-N-Sz}
H\ket{\nu; N,S^{z}} =E_\nu(N,S^{z})\ket{\nu;N,S^{z}},
\end{equation}
where the index $\nu$ labels  a particular state in the sector with fixed eigenvalues of $N$ and $S^z$. Due to the constraint of no doubly occupied sites, the   number of electrons obeys $N\leq L$, where $L$ is the number of sites. Note that for $N=L$ the Hamiltonian in Eq. (\ref{eq:model}) vanishes identically, and the system becomes a Mott insulator with a highly degenerate ground state, equivalent to the atomic limit of the Hubbard model at half-filling.

The pair hopping Hamiltonian also separately conserves  the number of electrons in distinct sub-lattices, namely, $N_A = \sum_{j\in A} n_{j}$ and $N_B = \sum_{j\in B} n_{j}$, where $A$ and $B$ denote the set of odd and even sites, respectively.
We are able to explicitly show that $[N_A,H]=[N_B,H]=0$ by using the identity
\begin{equation}\label{eq:commutator-n-b}
[n_{i},b_{j,j+1}]=-\left(\delta_{i,j}+\delta_{i,j+1}\right)b_{j,j+1}.
\end{equation}

However, to make contact with fracton physics, we   need to show that our Hamiltonian obeys a dipole conservation law.
Here it may seem that we run into trouble since the pair hopping Hamiltonian (\ref{eq:model}) does not conserve the ordinary dipole moment, $\sum_{j} j n_{j}$.
The solution to this issue was provided by Sous and Pretko \cite{Sous2020PRB}, and lies on the use of the staggered charge density $\tilde{n}_{j} = n_j e^{i\pi \sum_{l<j}n_{l}}$ to define the dipole operator
\begin{equation}\label{eq:dipole-operator}
D = \sum_{j} j \tilde{n}_{j}.
\end{equation}
The reason behind this choice is quite simple to check.
Since the Hamiltonian (\ref{eq:model}) only moves tightly  bound pairs, the magnitude of the dipole in Eq. (\ref{eq:dipole-operator}) is always left unchanged so $[D,H]=0$.
By the same token, the ordinary hopping of a  single electron would change the value of $D$ and spoil the conservation law.

\begin{figure}
\includegraphics[scale=0.7]{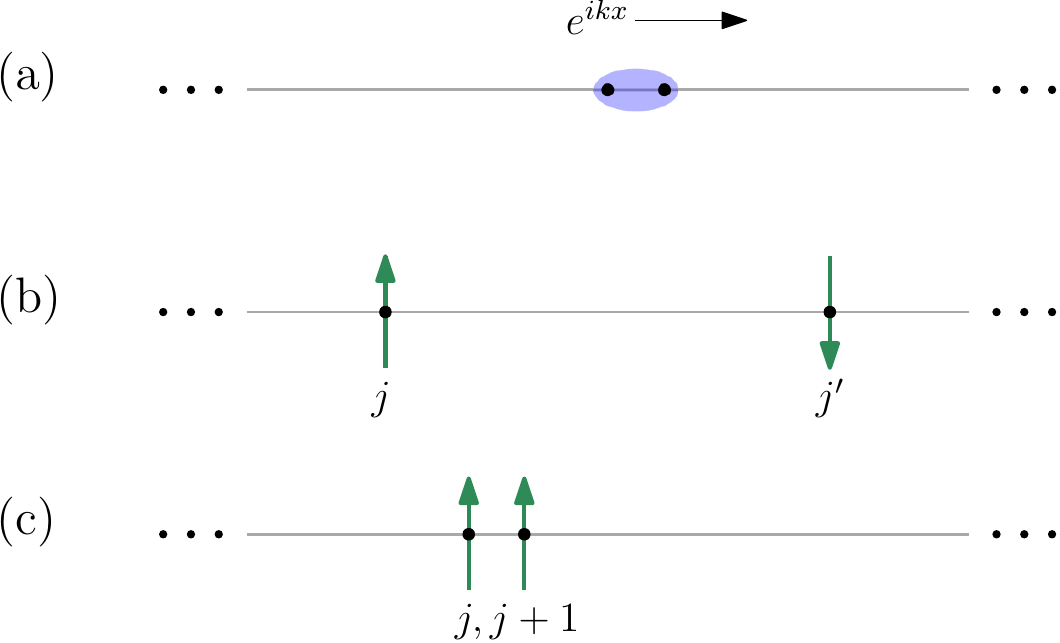}
\caption{Three distinct types of particles in the two-electron sector: (a)  itinerant singlet pair; (b)  two isolated electrons; and (c)  localized  triplet pair.}
\label{fig:2e_eigenstates}
\end{figure}

The minimum number of electrons for which the system has some dynamics is $N=2$. In this sector, the ground state has two electrons bound into a singlet pair, which can then occupy the state with lowest kinetic energy. There are two ways to break this singlet pair: we can either separate the two electrons by at least one lattice site or change the spin state of two nearest-neighbor electrons into a triplet  (see Fig. \ref{fig:2e_eigenstates}).

This observation implies that the separate numbers of nearest-neighbor pairs and single electrons can be used to further characterize the system for $N>2$.
We then introduce the notation $(M_{0},M_{1/2},M_{1})$ to designate a subspace with  $M_0$ singlet pairs, $M_{1/2}$ single electrons, and $M_1$ triplet pairs. 
Clearly, these three  occupation numbers must satisfy the selection rule $N=2M_{0}+M_{1/2}+2M_{1}$.
Strictly speaking,  this three-component classification is  valid only asymptotically    as it    refers to the local basis for well-separated particles.
In fact, when a singlet pair and a single electron occupy three adjacent sites, this occupation number  basis   is complete but not orthonormal, as we will discuss in more detail in Sec. \ref{sec:singleparticle}.
Formally, in the many-body problem the occupation numbers $(M_{0},M_{1/2},M_{1})$ label sets of Krylov subspaces \cite{Serbyn2020} of states connected by repeated action of the Hamiltonian, constructed starting from a product state where the different particles are separated and the counting can be performed. This classification will prove useful when studying low-lying excitations above the ground state.


\section{Ground state as a liquid of singlet pairs}\label{sec:ground-state}

For even values of $N\geq 2$, the ground state is in the subspace $(N/2,0,0)$, where all electrons form   singlet pairs which can gain kinetic energy.
We will prove in Sec. \ref{sec:excitations} that both spin-triplet  and single-electron excitations are gapped below the half-filled, insulating regime.
In this section, we present a coordinate Bethe ansatz solution to the $(N/2,0,0)$ subspace that allows us to solve the ground state and deduce the corresponding low-energy theory. We show that the spectrum contains gapless excitations corresponding to adding or removing singlet pairs. 
Our exact solution cements previous results \cite{Batista1995} while bringing out an amusing connection with exclusion models \cite{Gomez-Santos1993, Dias2000}.


\subsection{Bethe ansatz solution}

We represent the quantum state vector of $M$ singlet pairs in the following form
\begin{equation}\label{eq:M00-state}
\ket{\Psi} = \sum_{x_1,\dots,x_M} \Psi(x_{1},\dots,x_{M}) \prod_{j=1}^Mb_{x_{j}-1/2,x_j+1/2}^{\dagger}  \ket{0},
\end{equation}
where $\ket{0}$ is the vacuum  state.
The half-integer variables $x_1,\dots,x_M$ are center-of-mass coordinates that label the singlet bonds.
We take them to be ordered as $x_{1}<x_{2}<\dots<x_{M}$ since other arrangements follow from symmetry of the wave function with respect to permutation of two singlet pairs.
Moreover,  the no-double-occupancy constraint for electrons imposes a no-nearest-neighbor condition for  singlet pairs, e.g., 
\begin{equation}
\mathcal{P}b_{j,j+1}^{\dagger}b_{j+1,j+2}^{\dagger} \ket{0} =0.
\end{equation}
Thus, we require the singlet bond positions to satisfy $x_{i+1}> x_{i}+1$, for $i=1,\dots,M-1$.

At this point we compactify our linear chain in Eq. (\ref{eq:model}) into a ring with $L$ sites by imposing periodic boundary conditions.
We then assume the following ansatz for the wave function:
\begin{equation}\label{eq:M00-BA-wave-function}
\Psi(x_{1},\dots,x_{M}) =\sum_{P} A(P)\exp\Bigg(i\sum_{j=1}^{M}k_{Pj}x_{j}\Bigg),
\end{equation}
where the summation runs over all permutations  $P$  of the quasimomenta $(k_{1},\dots,k_{M})$ that specify the Bethe state. 
The energy $E$ and momentum $Q$ of the  state are set, respectively, by the kinetic energy of well separated pairs and by the eigenvalue upon translation by one lattice site.
We have
\begin{equation}\label{eq:energy-momentum-M00}
E[k]=-2\sum_{j=1}^{M}\cos k_{j},\qquad Q[k]=\sum_{j=1}^{M}k_{j},
\end{equation}
where we set the lattice spacing to unity. 
The amplitudes $A(P)$, on the other hand, are fixed by the two-body scattering amplitude. 
When two singlet pairs scatter, they exchange their quasimomenta.
If two  permutations $P$ and $P'$ differ only by a pair of quasimomenta such that that $k_{Pj}=k_{P'j+1}=k$ and $k_{Pj+1}=k_{P'j}=k'$, the matching conditions imply  $A(P)e^{ik'}+A(P')e^{ik}=0$.
As a result,  we find that the singlet pair scattering matrix $S(k,k')\equiv A(P')/A(P)$ is given by 
\begin{equation}\label{eq:singlet-pair-S-matrix}
S(k,k')=-e^{-i(k-k')}.
\end{equation}

The periodicity of the wave function $\Psi(x_{2},\dots,x_{M},x_{1}+L)=\Psi(x_{1},x_{2},\dots,x_{M})$ imposes a set of $M$ quantization conditions for the $M$ quasimomenta, known as Bethe equations,
\begin{equation}
e^{ik_{j}L} \prod_{l\neq j}^{M} S(k_{j},k_{l})=1,\qquad j=1,2,\dots,M.
\end{equation}
Taking the logarithm  of these equations, we   arrive at
\begin{equation}\label{eq:M00-Bethe-eq}
k_{j}(L-M+1)+\sum_{l\neq j}k_{l}=2\pi I_{j}.
\end{equation}
Here the branch of the logarithm  is parametrized by an integer $I_{j}$ for odd $M$, or by a half-integer $I_{j}$ for even $M$.
Alternatively, one can replace the sum over quasimomenta in Eq. (\ref{eq:M00-Bethe-eq}) by the total momentum $Q[k]$, obtaining
\begin{equation}
k_{j}(L-M)+Q[k]=2\pi I_{j}.
\end{equation}
The equation above makes touch with the quantization relation of exclusion models \cite{Gomez-Santos1993,Dias2000}. 
In such models, the extended hard-core condition reduces the effective size of the chain by a factor proportional to the number of particles. This simple fact gives  rise to Luttinger liquid behavior with a density-dependent Luttinger parameter.


\subsection{Ground state properties}

\begin{figure}[t]
\includegraphics[width=\columnwidth]{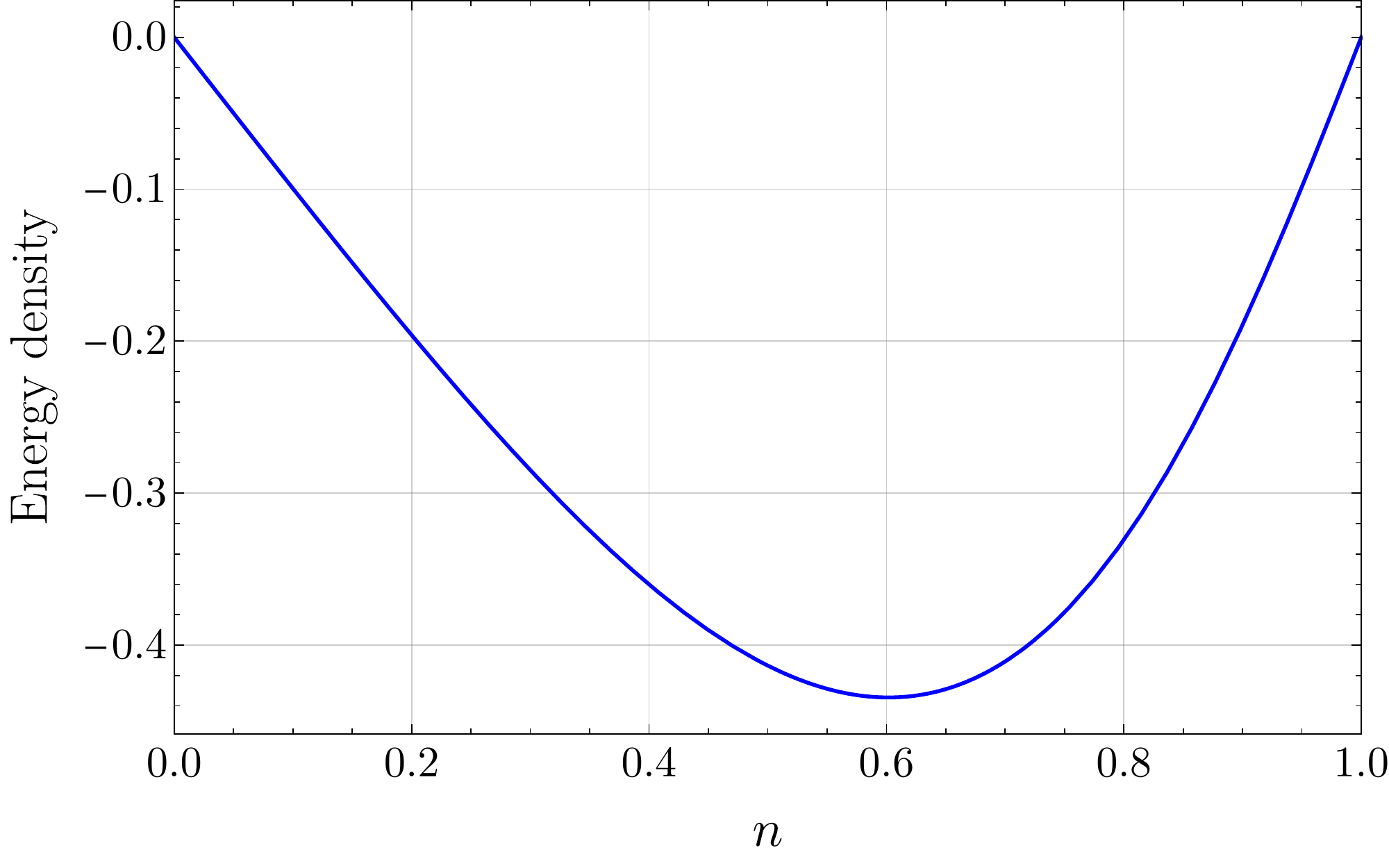}
\caption{Ground state energy density as a function of the electronic filling $n$.}
\label{fig:GS_energy_density}
\end{figure}

The ground state  for $M=N/2$ singlet pairs corresponds to picking  the mode numbers
\begin{equation}
I_{j}=\left\{ -\frac{M-1}{2},-\frac{M-3}{2},\dots,\frac{M-1}{2}\right\} .
\end{equation}
Notice that this choice already takes into account the fact that $I_{j}$ is integer or half-integer depending on  the parity of  $M$.
Substituting  the quasimomenta   back into Eq. (\ref{eq:energy-momentum-M00}), we find that  the ground state has zero momentum and   energy
\begin{equation}\label{eq:GS-energy-M00}
E_{g}(N)=-2\csc\left(\frac{2\pi}{2L-N}\right)\sin\left(\frac{\pi N}{2L-N}\right).
\end{equation}
If  we now take the thermodynamic limit, $L\rightarrow\infty$ with fixed $n\equiv N/L$, we obtain the ground state energy density
\begin{equation}\label{eq:GS-energy-density-M00}
E_{g}/L=-\left(\frac{2-n}{\pi}\right)\sin\left(\frac{\pi n}{2-n}\right).
\end{equation}
Figure \ref{fig:GS_energy_density} shows the dependence of the ground state energy on the electron density. We see that, unlike the exclusion model of spinless fermions   \cite{Dias2000}, the energy minimum for the singlet-pair liquid occurs at an incommensurate filling $n\simeq0.602$.

Having found  the solution for $(M,0,0)$ subspaces,  we are now in position to compute the two-particle excitation gap.
The latter is defined as the difference between the chemical potentials to add and to remove two particles from the system.
Namely,
\begin{equation}
\Delta_{2p}=E_{g}(N+2)-2E_{g}(N)+E_{g}(N-2),
\end{equation}
where $E_{g}(N)$ denotes the ground state energy for $N$ electrons. 
Evaluating this gap in the large system size limit ($L\gg 1$) yields
\begin{equation}
\Delta_{2p}=\frac{\pi}{L}\frac{16}{\left(2-n\right)^{3}}\sin\left(\frac{\pi n}{2-n}\right)+O(1/L^{2}).\label{eq:2p-gap}
\end{equation}
This shows that  the two-particle gap scales as $\Delta_{2p}\sim 1/L$   and vanishes in the thermodynamic limit.

We can also capture the effects of an external magnetic flux $\Phi$ threading the ring by  modifying  the Hamiltonian in Eq. (\ref{eq:model}) to
\begin{equation}\label{eq:model-flux}
H=-\sum_{j=1}^{L}\mathcal{P}\left(b_{j-1,j}^{\dagger}b_{j,j+1}e^{2i\theta}+\text{H.c.}\right)\mathcal{P},
\end{equation}
where $\theta=\Phi/L$ is the Peierls phase. Note the factor of 2 in the phase, associated with the charge $2e$ of the singlet pairs. 
Since the Hamiltonian remains translation invariant, we can repeat   the previous steps in the derivation of the Bethe equations.
This lead us to the following modification in the ground state energy density:
\begin{equation}\label{eq:GS-energy-M00-theta}
E_{g}/L=-\left(\frac{2-n}{\pi}\right)\sin\left(\frac{\pi n}{2-n}-2\theta\right).
\end{equation}
In the presence of a magnetic flux, the ground state acquires  a nonzero expectation value of the current operator
\begin{equation}
J= -2i\sum_{j=1}^{L}\mathcal{P}\left(b_{j-1,j}^{\dagger}b^{\phantom\dagger}_{j,j+1} - b_{j,j+1}^{\dagger}b^{\phantom\dagger}_{j-1,j}\right)\mathcal{P},
\end{equation}
defined from Eq. (\ref{eq:model-flux}) according to $J=(\partial H/\partial\theta)_{\theta=0}$. It follows that the singlet pair liquid shows metallic behavior in the sense of  nonvanishing charge transport at low energies. 


\subsection{Low-energy theory}

The vanishing of    the two-particle gap,   $\Delta_{2p}\sim 1/L$ for $L\to \infty$,  suggests that the  low-energy physics of the model corresponds to a Luttinger liquid of singlet pairs. 
Quite generally, the low-energy spectrum of interacting one-dimensional systems in the Luttinger liquid universality class is described by the effective Hamiltonian \cite{Haldane1981,Giamarchi2004}\begin{equation}
H_{\rm LL}=\sum_{q\neq0}v_S|q|a^\dagger_q a^{\phantom\dagger}_q +\frac{\pi v_N}{2L}(\Delta \hat N)^2+\frac{\pi v_J}{2L}\hat J^2.
\end{equation}
Here $a_q$ annihilates a bosonic mode, with quantized momentum $q=2\pi m/L$ with $m\geq 1$ for periodic boundary conditions,  that propagates with the sound velocity $v_S$.  The  operators $\Delta \hat N$ and $\hat J$ count the number of charge and current excitations, respectively, with associated velocity parameters $v_N$ and $v_J$. In our case, the gapless bosonic modes must be identified with collective charge fluctuations of the singlet-pair liquid. In Sec. \ref{sec:excitations} we will show that charge-$e$ and spin excitations are gapped. In this sense, the effective theory is analogous  to a Luther-Emery liquid \cite{Luther1974}.  However, in a conventional Luther-Emery liquid the gapped modes are described as mobile kinks in  a sine-Gordon model. In Sec. \ref{sec:localization}, we shall see that  the single-electron excitations of the singlet-pair liquid  depart from this behavior in  that  they have localized wave functions as a consequence of the dipole moment conservation law.

The velocities that characterize the low-lying excitations in the Luttinger liquid can be   obtained from our previous equations. They are given by 
\begin{align}
v_{N} & =\frac{L}{\pi}\left(\frac{\partial^{2}E_{g}}{\partial N^{2}}\right)=\frac{4}{(2-n)^{3}}\sin\left(\frac{\pi n}{2-n}\right), \nonumber \\
v_{J} & =\frac{\pi}{4L}\left(\frac{\partial^{2}E_{g}}{\partial\theta^{2}}\right)=(2-n)\sin\left(\frac{\pi n}{2-n}\right), \\
v_{S} & =\sqrt{v_{N}v_{J}}=\frac{2}{2-n}\sin\left(\frac{\pi n}{2-n}\right).\nonumber
\end{align}
Note that in order to use Eq. (\ref{eq:GS-energy-M00-theta}) we need to introduce a factor of four in the expression for $v_J$ since $\hat J$ is defined with respect to single electron excitations. Our results are  similar to those found in the study of exclusion models \cite{Gomez-Santos1993,Dias2000}, with important differences arising from the composite nature of the singlet pairs. Using the relation $K_{\rho}^{2}=v_{J}/v_{N}$, we can also obtain the Luttinger parameter $K_\rho = \frac{1}{2}(2-n)^2$, which determines the exponents in the power-law decay of correlation functions for spin-0 operators.  For instance, in the continuum limit the pair annihilation operator is represented by the bosonized form  $b_{j,j+1}\sim e^{-i\sqrt{2\pi/K_\rho}\Theta(j)}$, where\begin{equation}
\Theta(x)=\sum_{q\neq 0}\frac1{\sqrt{2|q|L}}\left(a^{\phantom\dagger}_{q}e^{iqx}+a^{\dagger}_{q}e^{-iqx}\right).
\end{equation}
As a result, the singlet-pair Green's function decays at large distances as  $\langle b_{0,1}^{\dagger}b^{\phantom\dagger}_{r,r+1}\rangle \sim r^{-1/K_{\rho}}$.

The Luttinger parameter characterizes the effective interactions in the liquid.  For low electronic densities,   $n< 2-\sqrt{2}\simeq 0.59$, we get $K_\rho>1$. In  a conventional  fermionic Luttinger liquid, as obtained for instance by bosonizing the Hubbard model \cite{Giamarchi2004}, one obtains  $K_\rho>1$ in the regime of attractive electron-electron interactions. In this case, superconducting correlations are dominant in the sense that they decay more slowly than other correlations. As pointed out by Batista \textit{et al.} \cite{Batista1995}, in the dilute limit  the physics of the singlet-pair liquid fits the simple picture of  a system of   hardcore bosons valid  in the strong coupling limit of  the \emph{attractive} Hubbard model. On the other hand, for $n>2-\sqrt{2}$, we obtain $K_\rho<1$,  characteristic of a fermionic system  with repulsive interactions.   In particular,  $K_\rho\rightarrow1/2$ as we approach half-filling, $n\to1$. This regime is dominated by ``charge-density-wave'' correlations, related to the staggered part of the density operator for the particles in the liquid. Interestingly, the density operator for singlet pairs corresponds to the Heisenberg operator:\begin{equation}
b^\dagger_{j,j+1}b^{\phantom\dagger}_{j,j+1}=\frac14n_{j}n_{j+1}-\mathbf S_j\cdot \mathbf S_{j+1},\label{Heisenberg}
\end{equation}
where $\mathbf S_j=\frac12\sum_{\alpha\beta}c^\dagger_{j\alpha}(\boldsymbol\sigma)_{\alpha\beta}c^{\phantom\dagger}_{j\beta}$ is the local spin operator. The staggered part of the two-spin operator in Eq. (\ref{Heisenberg}) is the order parameter for spin dimerization \cite{Majumdar1969,White1996}. This  tendency to dimerization (without true long-range order) is reminiscent of doped valence bond crystals \cite{Rokhsar1988,Poilblanc2008}.

\begin{figure}[t]
\includegraphics[width=\columnwidth]{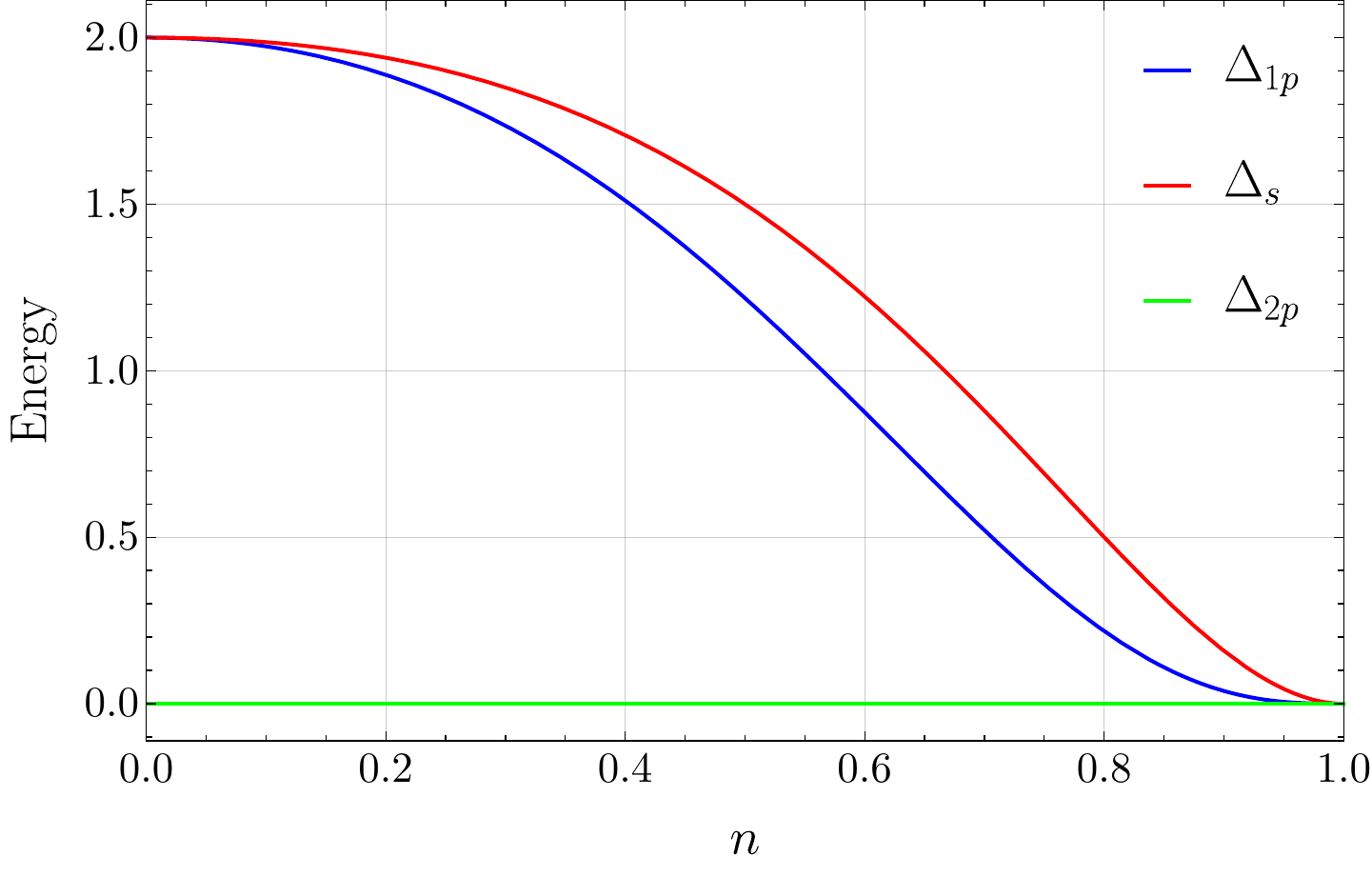}
\caption{One-particle ($\Delta_{1p}$), two-particle ($\Delta_{2p}$), and spin ($\Delta_{s}$) excitation gaps as a function of the electronic density $n$.}
\label{fig:all_gaps}
\end{figure}


\section{Spin and single particle excitations}\label{sec:excitations}

In this section  we derive the energy gaps for spin-triplet  and single-electron excitations.
A quick summary of our results is available in Fig. \ref{fig:all_gaps}, where we plot the energy gaps in the thermodynamic limit.


\subsection{Spin excitations}

We are now interested in   the spin excitations above the singlet-pair ground state.
To this end we need to inspect the $(M,0,1)$ sectors of the model, which correspond to the subspaces with an arbitrary number of singlet pairs and just one nearest-neighbor triplet pair. Given  the SU(2) spin-rotation symmetry of the model, we can choose to analyze the triplet pair with two spin-up electrons, {\it i.e.}, the $S^z=1$ sector.

The quantum state vector for $M$ singlet pairs and one triplet pair with $S^z=1$ takes the form 
\begin{eqnarray}
\ket{\Psi} &=& \sum_{x_0}\sum_{x_1,\dots,x_M} \Psi(x_{0};x_{1},\dots,x_{M}) c_{x_{0}-1/2,\uparrow}^{\dagger}c_{x_{0}+1/2,\uparrow}^{\dagger}\nonumber\\
&&\times b_{x_{1}-1/2,x_1+1/2}^{\dagger} \dots  b_{x_{M}-1/2,x_M+1/2}^{\dagger} \ket{0} .
\end{eqnarray}
The bond coordinates $x_0$ for the triplet  and $x_1,\dots,x_M$ for the singlet pairs run through half-integer values, and we will assume that $ r_{1}<r_{2}<\dots<r_{M}$ for $r_i=x_i-x_0 $. To find the solution, we first separate the  motion of the singlet pairs relative to the position of the triplet pair.
That is, we write
\begin{equation}
\Psi(x_{0};x_{1},\dots,x_{M})=e^{iQx_{0}}f(x_{1}-x_{0},\dots,x_{M}-x_{0}),
\end{equation}
where $Q$ is the total momentum and 
\begin{equation}\label{eq:BA-box}
f(r_{1},\dots,r_{M})=\sum_{P} A(P)\exp\Bigg( i \sum_{j=1}^{M} k_{Pj}r_{j}\Bigg).
\end{equation}
Here, differently from the ansatz employed in Sec. \ref{sec:ground-state}, the sum   must run over all permutations and also negations of $(k_1,\dots,k_M)$ \cite{Zvyagin2005}.

Substituting the state into the eigenvalue equation $H\ket{\psi} = E\ket{\psi}$, we find that the energy of such state is set by the kinetic energy of well separated pairs as in Eq. (\ref{eq:energy-momentum-M00}). 
The scattering matrix between two singlet pairs is still the same as Eq. (\ref{eq:singlet-pair-S-matrix}). 
The difference from Sec. \ref{sec:ground-state} is that now the triplet sits at the boundaries of the interval, so the wave function  must satisfy
\begin{equation}
f(1,r_{2},\dots,r_{M})=f(r_{1},r_{2},\dots,L-1)=0.
\end{equation}
Here we see that the interaction with the triplet pair simply enforces hard wall boundary conditions for the singlet pairs.
These equations imply that the $k$ modes are perfectly reflected at the boundaries of the interval. 
Thus, if two permutations $P$ and $P'$ differ only by $k_{P1}=-k_{P'1}=k$,   the boundary condition at $r_1=1$ gives
\begin{equation}\label{eq:scatter-reflection-left}
s_{1}(-k,k)\equiv A(P)/A(P') = -e^{-2ik}.
\end{equation}
Likewise, if  $P$ and $P'$ differ only by $k_{PM}=-k_{P'M}=k$, we obtain from the   boundary at $r_M=L-1$
\begin{equation}\label{eq:scatter-reflection-right}
s_{L-1}(k,-k) \equiv A(P')/A(P) = -e^{2ik(L-1)}.
\end{equation}

\begin{figure}[t]
\includegraphics[scale=.7]{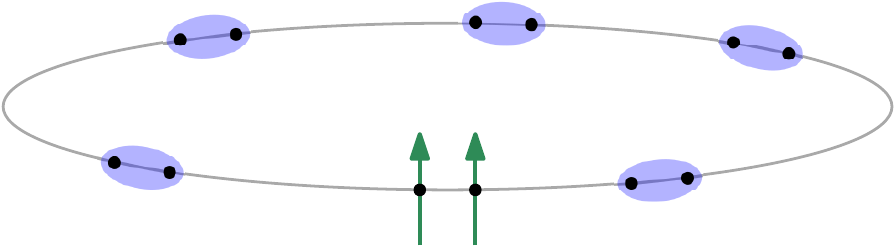}
\caption{Spin-triplet-pair excitation in the singlet-pair liquid. The localized  triplet acts as a hard wall for itinerant singlet pairs.}
\label{fig:spin-excitation}
\end{figure}

The $M$ quasimomenta $k_j$ are determined by solving the eigenvalue equation $T_{j}= 1$, where $T_{j}$ is the translation operator that swipes the entire interval:
\begin{eqnarray}
T_{j} & =&S(k_{j},k_{j+1})\dots S(k_{j},k_{M}) s_{L-1}(k_j,-k_j)\nonumber\\
&&\times S(k_{M},-k_{j})\dots S(k_{j+1},-k_{j})S(k_{j-1},-k_{j})\dots\nonumber\\
 && \times  S(k_{1},-k_{j})s_{1}(-k_j,k_j)S(k_{1},k_{j})\dots S(k_{j-1},k_{j}).\nonumber\\
\end{eqnarray}
Hence, by using that $S(k,q)S(q,-k)=e^{-2ik}$, the quantization equation is simplified to $e^{2ik_{j}(L-M-1)}=1$.
Taking the logarithm  of this expression, we arrive at the desired Bethe equations
\begin{equation}\label{eq:M01-quant-cond}
k_{j}(L-M-1)=\pi I_{j},
\end{equation}
where $I_j$ is an integer with  $1\leq I_j\leq L-M-2$.

This solution shows that triplet pairs just play the role of hard walls for the itinerant singlet pairs (see Fig. \ref{fig:spin-excitation}). Therefore, their sole effect is to change boundary conditions.
The energy levels on the ring are degenerate with respect to the total momentum $Q$, or equivalently, to the position of the triplet pair.
The lowest energy configuration in the $(M,0,1)$ subspace corresponds to picking  the mode numbers
\begin{equation}
I_{j}=\{1,2,\dots,M\}.
\end{equation}
We then compute the energy by substituting the corresponding quasimomenta into   Eq. (\ref{eq:energy-momentum-M00}).
The result is
\begin{align}\label{eq:GS-energy-M01}
&E_{g}(N,S^z=1) \nonumber\\
&\hspace{1cm}=1-\csc\left(\frac{\pi}{2L-N}\right)\sin\left[\frac{\pi(N-1)}{2L-N}\right],
\end{align}
where $N=2M+2$ is the total number of electrons, and $E_{g}(N,S^{z})$ designates the ground state energy for $N$ electrons and total spin $z$-polarization $S^{z}$.

The spin gap is defined as the excitation energy from the singlet ground state to the lowest-lying triplet state,
\begin{equation}
\Delta_{s}=E_{g}(N,S^{z}=1)-E_{g}(N,S^{z}=0).
\end{equation}
Using Eqs. (\ref{eq:GS-energy-M00}) and (\ref{eq:GS-energy-M01}), we find 
\begin{equation}\label{eq:spin-gap}
\Delta_{s}=1+\cos\left(\frac{\pi n}{2-n}\right)+O(1/L),
\end{equation}
in the large system size limit, $L\gg 1$. As shown in Fig. \ref{fig:all_gaps}, the spin gap $\Delta_{s}$  only closes at the insulating point $n=1$. For $n\to1$, the spin gap displays a quadratic dependence on the deviation from half filling, 
\begin{equation}
\Delta_{s}=2\pi^{2}(1-n)^{2}+O(1-n)^{3}.
\end{equation}
Our results agree with Batista \textit{et al.}  \cite{Batista1995}, who found this energy gap by adding a triplet in the middle of an open chain   with a finite density of singlet pairs.

The case with more triplet pairs can be treated in similar fashion.
In such subspaces, the infinitely heavy triplet pairs create an effective disordered landscape for the singlets and we need to consider several partitions on the ring.
This situation resembles  a quantum disentangled liquid \cite{Grover2014,Smith2017a}, a fluid made out of  two species of particles with a large mass ratio. We also note that the hard-wall nature of the triplet pair generalizes to larger clusters of $m>2$ electrons occupying neighboring sites with maximum total spin   $S=m/2$.  The reason is that  the singlet-pair hopping  Hamiltonian cannot generate any dynamics when applied to a state which is completely symmetrized with respect to the spin degree of freedom. Our triplet pairs are analogous to the frozen states discussed for the spin-1 chain model in Ref. \cite{Sala2020}.


\subsection{Single-electron excitations}\label{sec:singleparticle}

We now  consider the $(M,1,0)$ subspace, in which   all electrons but one are bound into singlet pairs. 
However, in contrast with the other subspaces considered so far, the $(M,1,0)$ subspace is not amenable to Bethe ansatz and we have not found general solutions in this case.
Without a general solution, we begin by treating the case with only one singlet pair in addition to the single electron, {\it i.e.}, we first consider the $(1,1,0)$ subspace. 
Despite obvious limitations, this exact solution will serve   as a valuable source of insight into the many-body problem.
Fortuitously, it will be enough to identify the lowest-energy configuration that allow us to compute the associated energy   gap for an open chain at finite  density of singlet pairs.

\begin{figure}[t]
\includegraphics[scale=0.75]{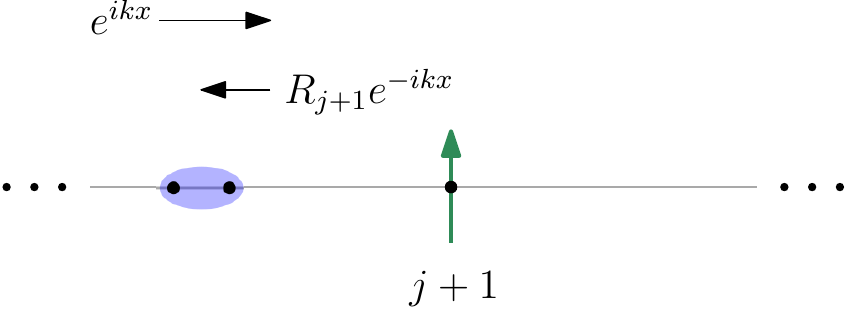}\vspace{.5cm}
\includegraphics[scale=0.75]{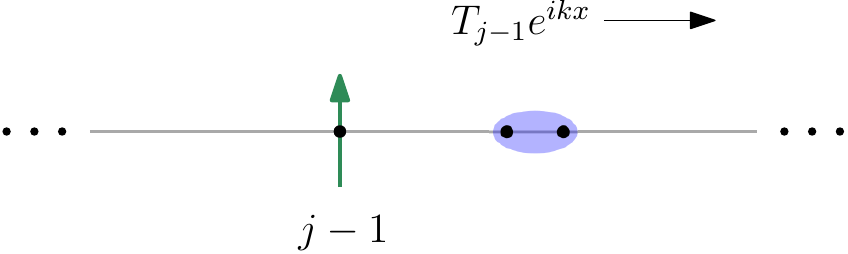}
\caption{Scattering state with a right-moving singlet pair. From top to bottom, the  electron jumps two sites to the left  when the singlet pair is transmitted.}
\label{fig:right-scattering}
\end{figure}

In the $(1,1,0)$ subspace, the quantum states can be written as
\begin{equation}\label{eq:110-quantum-state}
\ket{\psi} = \sum_{j,x} \psi(j;x)c_{j\sigma}^{\dagger} b_{x}^{\dagger}\ket{0}.
\end{equation}
The basis consists of  states specified by the position $j$ of the single electron and $x$ of the singlet bond, with ${|x-j|}> 1/2$.
The eigenfunctions $\psi(j;x)$  of the Hamiltonian in Eq. (\ref{eq:model}) 
must obey the   lattice Schr\"odinger equation.
First, for $x<j-3/2$, only the singlet pair moves, and we have  
\begin{equation}\label{eq:110-singlet-hop}
E\psi(j;x)=-\psi(j;x-1)-\psi(j;x+1).
\end{equation}
Then, when the singlet pair and the electron meet, {\it i.e.}, for $|x-j|=3/2$, we get  
\begin{eqnarray}
E\psi(j;j-3/2)&=&-\psi(j;j-5/2)\nonumber\\ 
& &+\gamma\psi(j-2;j+1/2),\nonumber\\
E\psi(j;j+3/2)&=&-\psi(j;j+5/2)\label{eq:110-meet}\\ 
&& +\gamma\psi(j+2;j-1/2),\nonumber
\end{eqnarray}
where we have introduced a dimensionless constant $\gamma=1/2$ for notational convenience. Equation (\ref{eq:110-meet}) shows that the electron can move by two sites when  a singlet pair  is transmitted across it (see Fig. \ref{fig:right-scattering}). 
Finally, when $x>j+3/2$, the singlet pair is once again well separated from the electron and we recover Eq. (\ref{eq:110-singlet-hop}).

Before we move on, a cautionary remark is in order. As the reader may well have noticed, Eqs. (\ref{eq:110-meet}) are not Hermitian. However, this is nothing but an artifact of our basis choice (\ref{eq:110-quantum-state}), which is complete, but not orthonormal. In fact, this issue only appears when a pair and a single electron occupy three adjacent sites, as the inner product between the states $c_{j-1\sigma}^{\dagger}b_{j,j+1}^{\dagger}\ket{0} $ and $b_{j-1,j}^{\dagger}c_{j+1\sigma}^{\dagger}\ket{0} $ reveals:
\begin{equation}
\bra{0} c_{j+1\sigma} b_{j-1,j} c_{j-1\sigma}^{\dagger}b_{j,j+1}^{\dagger} \ket{0} =-1/2.
\end{equation}
A possible way to circumvent this matter is by adopting the following normalized orthogonal operators
\begin{align}
v_{j,1}^{\dagger}&= 
c_{j-1\sigma}^{\dagger}b_{j,j+1}^{\dagger} 
+ b_{j-1,j}^{\dagger}c_{j+1\sigma}^{\dagger},\nonumber\\
v_{j,2}^{\dagger}&=
\frac{1}{\sqrt{3}}\left(
c_{j-1\sigma}^{\dagger}b_{j,j+1}^{\dagger}
-b_{j-1,j}^{\dagger}c_{j+1\sigma}^{\dagger}
\right).
\end{align}
In terms of these operators, the quantum state (\ref{eq:110-quantum-state}) may be recast as
\begin{eqnarray}
\ket{\psi'} &=&\sum_{j} \left[\sum_{x<j-3/2}\psi(j;x)c_{j\sigma}^{\dagger}b_{x}^{\dagger}+\upsilon_{1}(j)v_{j,1}^{\dagger}\right.\nonumber\\
&&\left.+\upsilon_{2}(j)v_{j,2}^{\dagger}+\sum_{x>j+3/2}\psi(j;x)c_{j\sigma}^{\dagger}b_{x}^{\dagger} \right]\left|0\right\rangle ,
\end{eqnarray}
where the old non-orthonormal amplitudes are related to the new orthonormal ones according to
\begin{align}
\psi(j-1;j+1/2)&=\frac{1}{2}\upsilon_{1}(j)+\frac{\sqrt{3}}{2}\upsilon_{2}(j),\nonumber\\
\psi(j+1;j-1/2)&=\frac{1}{2}\upsilon_{1}(j)-\frac{\sqrt{3}}{2}\upsilon_{2}(j).
\end{align}
As a matter of fact, by employing these relations, one can make Eqs. (\ref{eq:110-meet}) manifestly Hermitian.

Up to this point we have not explored the  perks that come with the conservation of the dipole operator $D$ in Eq. (\ref{eq:dipole-operator}). To take full advantage of the latter,  we now turn to chains with open boundary conditions. In the infinite line, the solution takes the form of singlet pair scattering states. For clarity, we write right and left-moving components separately. For instance, the right-moving scattering state illustrated in Fig. \ref{fig:right-scattering} is given by
\begin{align}
\ket{\psi_{+}} &= 
\sum_{x<j}\Big(e^{ikx}+R_{j+1}e^{-ikx}\Big)c_{j+1,\sigma}^{\dagger}b_{x}^{\dagger} \ket{0}\nonumber\\
&\hspace{1cm}+\sum_{x>j}T_{j-1}e^{ikx}c_{j-1,\sigma}^{\dagger}b_{x}^{\dagger} \ket{0} .
\end{align}
Reflection and transmission amplitudes are fixed by Eqs. (\ref{eq:110-meet}) to be
\begin{align}
R_{j+1}&=-e^{ik(2j+1)}\left(\frac{1-\gamma^{2}}{1-\gamma^{2}e^{4ik}}\right),\nonumber\\
T_{j-1}&=-\gamma e^{-ik}\left(\frac{1-e^{4ik}}{1-\gamma^{2}e^{4ik}}\right).
\end{align}
As a check, note that these amplitudes satisfy the conservation of  probability, $|R_{j+1}|^{2}+|T_{j-1}|^{2}=1$. Likewise, the left-moving singlet takes the form
\begin{align}
\ket{\psi_{-}} &=
\sum_{x<j}\tilde T_{j+1}e^{-ikx}c_{j+1,\sigma}^{\dagger}b_{x}^{\dagger} \ket{0}\nonumber\\
&\hspace{.5cm}+\sum_{x>j}\Big(e^{-ikx}+\tilde R_{j-1}e^{ikx}\Big)c_{j+1,\sigma}^{\dagger}b_{x}^{\dagger} \ket{0} ,
\end{align}
where $\tilde R_{j-1} = R_{j+1} e^{-4ikj}$ and $\tilde T_{j+1}=T_{j-1}$. Thus, the most general form of  the solution   is $\ket{\psi}=A_{+}\ket{\psi_{+}}+A_{-}\ket{\psi_{-}}$, where $A_{\pm}$ are the corresponding amplitudes.

If we now put our system in  an open  chain with finite length $L$, the hard-wall boundary conditions   require that eigensolutions be a particular superposition of right- and left-moving scattering states. On the one hand, the condition  $\psi(j;1/2)=0$   yields   $A_{+}/A_{-}=-T_{j-1}/(e^{ik}+R_{j+1})$. On the other hand, $\psi(j;L+1/2)=0$ imposes the quasimomentum $k$ to be a solution of
\begin{align}\label{eq:110-quant-cond}
&\sin\left(kD\right)\sin\left[k(L+1-D)\right] \nonumber\\
&\hspace{1.5cm}=\gamma^{2}\sin\left[k(D-2)\right]\sin\left[k(L-1-D)\right],
\end{align}
where $D$ takes integer values from $2$ to $L-1$.

\begin{figure}[t]
\includegraphics[width=\columnwidth]{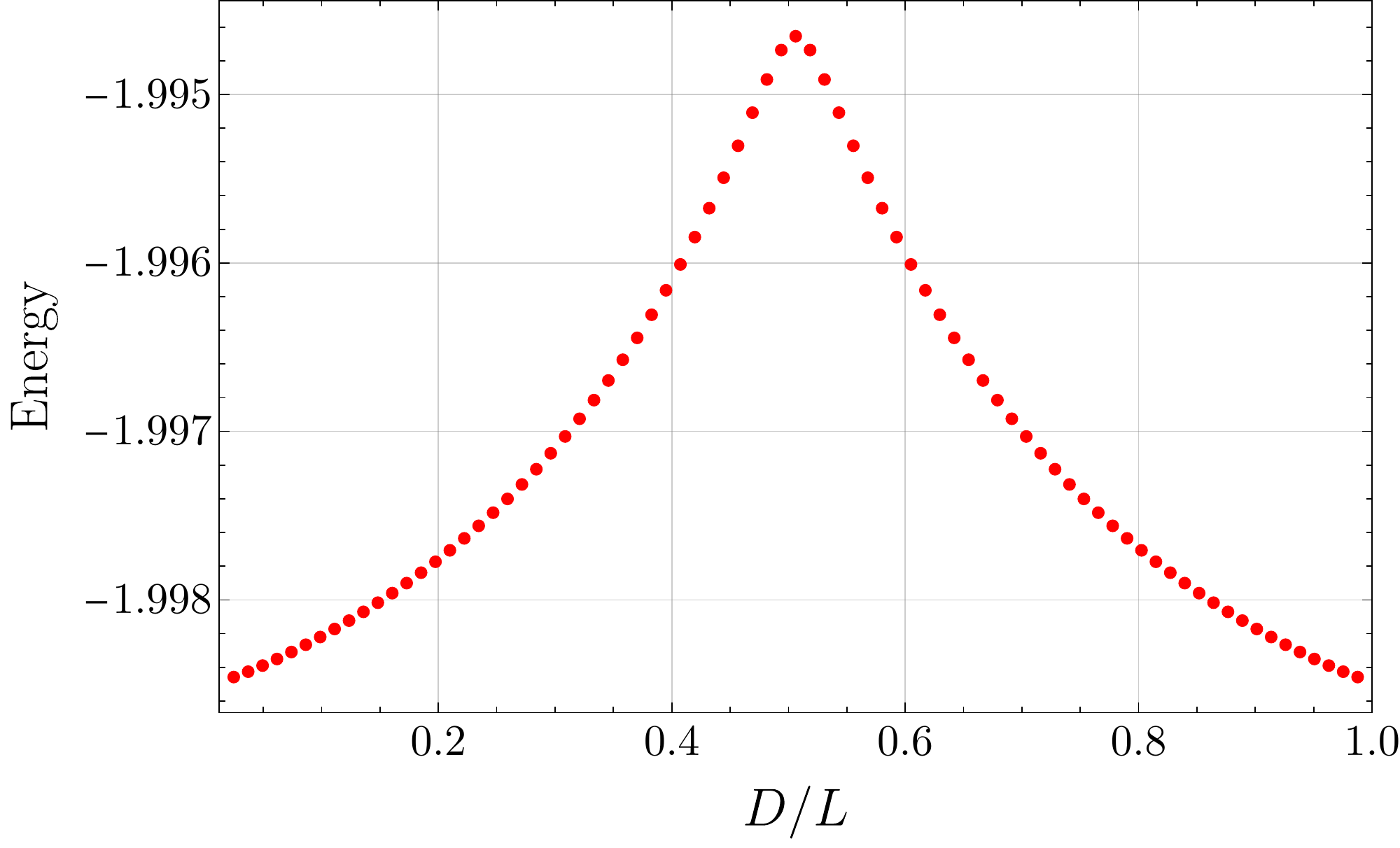}
\caption{Ground state energy in the $(1,1,0)$ subspace as a function of the dipole moment $D$. The numerical data was obtained for an open  chain  with $L=81$ sites.}
\label{fig:GS_energy_110_wire}
\end{figure}

\begin{figure}[t]
\includegraphics[width=\columnwidth]{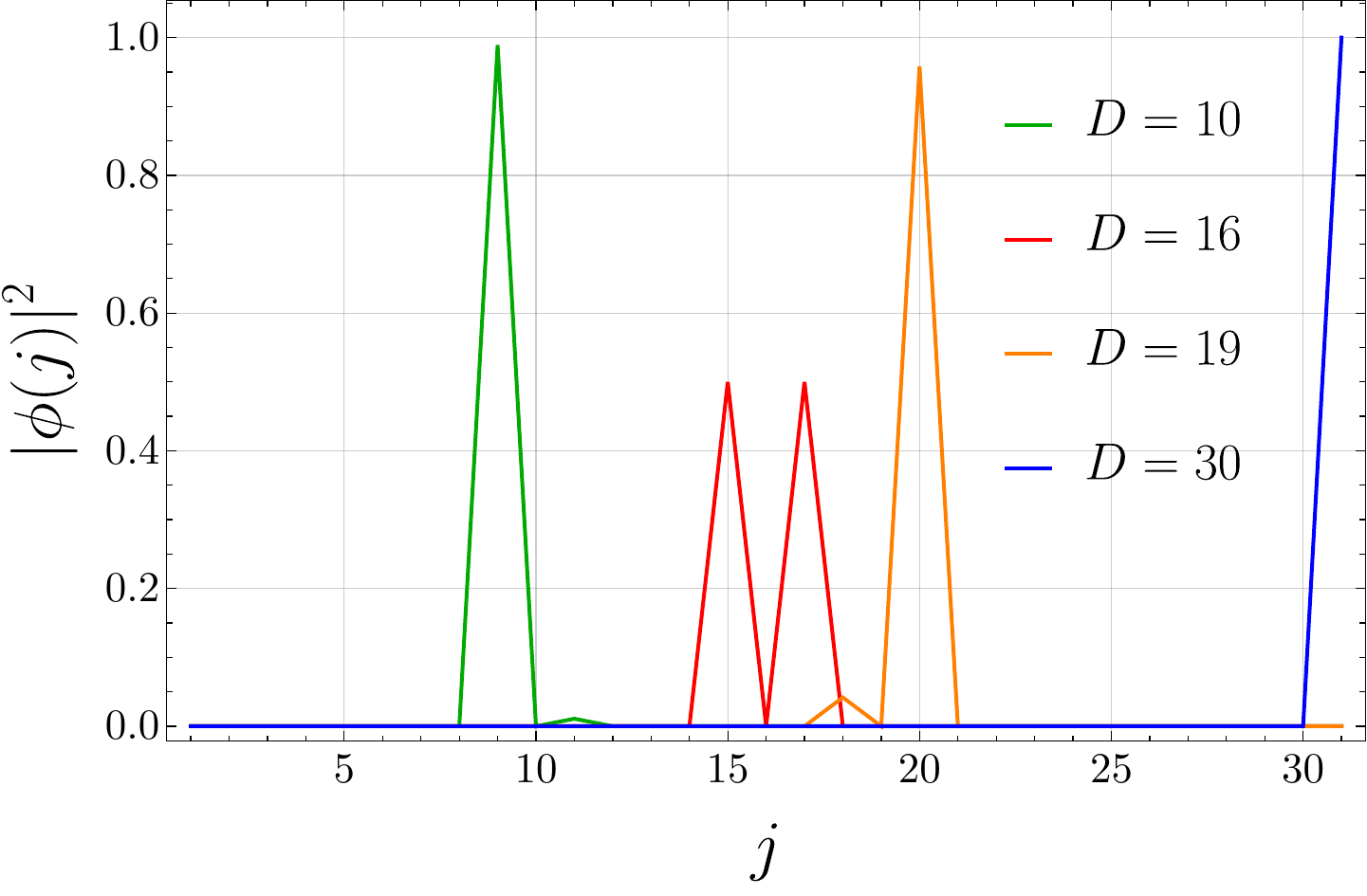}
\caption{Single-electron probability density in the $(1,1,0)$ space at various values of $D$. The numerical data was obtained for a wire with $L=31$ sites.}
\label{fig:ElectronWaveFunction-110}
\end{figure}

We can now solve Eq. (\ref{eq:110-quant-cond}) numerically for fixed values of the chain length $L$ and the conserved dipole moment $D$. The energy  of the scattering state is simply $E=-2\cos k$ for an allowed value of $k$.  
In Fig. \ref{fig:GS_energy_110_wire}, we plot the minimum energy  as a function of  $D$ for $L=81$.
The  minimum energy is obtained for   $D=2$ and $D=L-1$, corresponding to the electron sitting at one of the boundaries. The energy has a  peak for $D$ at the center of the chain.
(We note in passing that the  same behavior is observed  for an open chain in the subspace with one singlet pair and one triplet pair.)
We also plot in Fig. \ref{fig:ElectronWaveFunction-110} the electron probability density defined as  $|\phi(j)|^2\equiv \sum_x |\psi(j;x)|^2$. 
Note that in the (1,1,0) subspace the single-electron wave function can be nonzero only  at two sites, fixed by the value of $D$. By comparing the two nonzero amplitudes, we can see that the electron probability density bends towards the nearest boundary. The asymmetry increases as $D$ deviates from the center of the chain. When placed right at one of the boundaries, the electron becomes completely trapped at the boundary  site because the singlet pair can no longer be transmitted across it.  
Indeed, for $D=2$ or $D=L-1$, the right-hand side of Eq. (\ref{eq:110-quant-cond}) vanishes and the quantization condition becomes the one of a free itinerant singlet in a chain with $L-1$ sites.

Clearly, the single electron at the boundary remains locked even if we add an arbitrary number of singlet pairs in the chain.
This means that we can compute the energy of such $(M,1,0)$ state by solving the problem of $M$ pairs on a chain with $L-1$ sites. The result  can be read off from Eq. (\ref{eq:M01-quant-cond}), the quantization condition for $M$ pairs in a chain with $L-2$ sites.
With this piece of information we are able to determine the one-particle excitation gap $\Delta_{1p}$, defined as the difference between the first electron affinity and the first ionization energies \cite{Kuhner}, 
\begin{equation}
\Delta_{1p}=E_{g}(N+1)-2E_{g}(N)+E_{g}(N-1).
\end{equation}
Proceeding this way, we find  in the large $L$ limit
\begin{align}
&\Delta_{1p}=4\left(\frac{1-n}{2-n}\right)\cos\left(\frac{\pi n}{2-n}\right)\nonumber\\
&\hspace{2.5cm}+\frac{2}{\pi}\sin\left(\frac{\pi n}{2-n}\right)+O(1/L).
\end{align}
Therefore, there is a finite gap for single-electron excitations in the pair liquid for all $n<1$. This gap closes very smoothly with cubic behavior near $n=1$,
\begin{equation}
\Delta_{1p}=\frac{16\pi^{2}}{3}(1-n)^{3}+O(1-n)^{4}.
\end{equation}
This  behavior is illustrated in Fig. \ref{fig:all_gaps}.

\begin{figure}[t]
\includegraphics[scale=0.8]{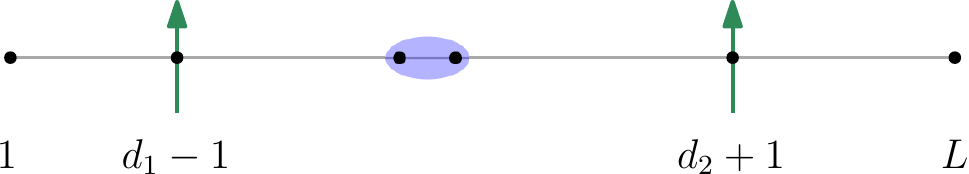}
\caption{Schematic representation  of an open  chain   with two   electrons and one singlet pair. The first and second  electrons can occupy the positions $d_1\pm1$ and $d_2\pm1$, respectively, depending on the position of the singlet pair. }
\label{fig:two-single-electrons}
\end{figure}

\begin{figure}[t]
\includegraphics[width=\columnwidth]{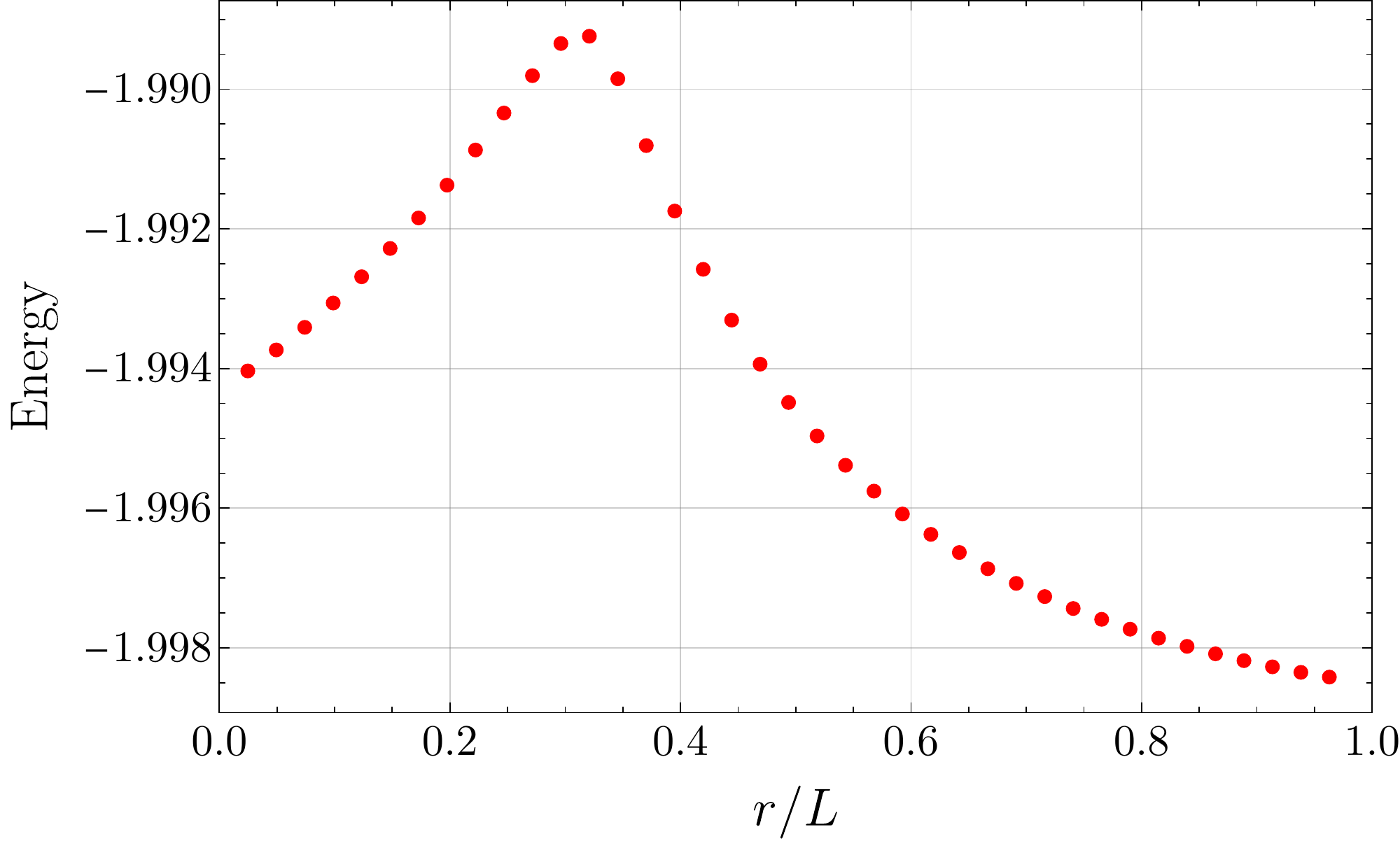}
\caption{Ground state energy as a function of the relative distance $r=d_2-d_1$ between the two   electrons in the presence of an itinerant singlet pair. The center of mass coordinate is  fixed at the center of the chain, $d_1+d_2=L+1$. Here we set $L=81$.}
\label{fig:GS-energy-120}
\end{figure}

We have also explored the exact solution for a system   with  one singlet pair and two well-separated electrons in an  open chain. In this case, we parametrize the scattering states by $d_1$ and $d_2$, which represent  the average positions of the electrons as shown in Fig. \ref{fig:two-single-electrons}. Note  that  the electrons can hop between  the sites $d_{1,2}\pm1$ as the singlet pair moves across them. The conserved dipole moment is $D=d_2-d_1-1$. In Fig. \ref{fig:GS-energy-120}, we show the lowest energy  for a chain with $L=81$ sites where we  fix  the ``center of mass" of the electrons to be  $(d_1+d_2)/2=(L+1)/2$ and vary the distance $r=d_2-d_1=D+1$. We see that for $r< L/3$ the single electrons feel a mutual attraction as the energy decreases with decreasing distance.  The interaction energy is of order $1/L$ in this (1,2,0) subspace.  Eventually, for large enough distance, $r> L/3$, the electrons become  closer to the boundaries than to each other, and the attraction to the boundary prevails. We interpret this result as a tendency of fractons to cluster together \cite{Prem2018,Pretko2020} by exchanging singlet pairs. However, it remains to be seen whether this attraction is manifest in the finite density regime.


\section{Localized electron wave functions}\label{sec:localization}


In this section we consider the wave function for a single electron immersed in the liquid with  an arbitrary number of singlet  pairs. While this sector is not integrable, we put forward an  approximation that  provides a simple physical picture for the electron as a type of ``dynamic boundary'' which moves every time  a singlet pair tunnels across it. Moreover,  the approximation  captures the localization of the electron wave function, a property protected by the conservation of the dipole moment.


\subsection{Constraint on single-electron position}

Let $\Psi(X;x_{1},\dots,x_{M})$ be the wave function of   $M$ singlet pairs and one single electron in an open chain, associated with the quantum state
\begin{eqnarray}
\ket{\Psi} &=& \sum_{X}\sum_{x_1,\dots,x_M} \Psi(X;x_{1},\dots,x_{M}) c_{X,\sigma}^{\dagger}\nonumber\\
&&\times b_{x_{1}-1/2,x_1+1/2}^{\dagger} \dots  b_{x_{M}-1/2,x_M+1/2}^{\dagger} \ket{0} .
\end{eqnarray}
We assume the singlet-pair coordinates  are ordered so that $x_{1}<\dots<x_{M}$.

\begin{figure}
\includegraphics[scale=0.7]{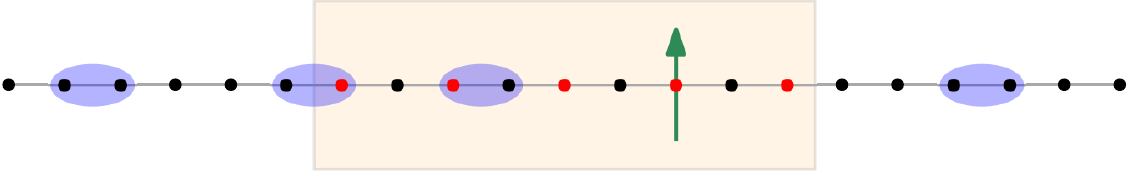}
\caption{Constrained dynamics of a single electron immersed in the singlet-pair liquid. Here we show a schematic picture of a state in the $(4,1,0)$ subspace. The  electron can only hop between the red sites inside  the shaded area.}
\label{fig:single-electron-excitation}
\end{figure}

Crucially,   the allowed values of $X$ for which the wave function is nonzero are constrained by the number of singlet pairs and the conservation of the dipole moment.  The dipole moment is given by \begin{equation}
D=X+M-2m,\label{DXMmrelation}
\end{equation}
where $m$ is the number of pairs to the left of the electron. In a sector with fixed $D$ and $M$, Eq. (\ref{DXMmrelation}) implies   $X=X(m)=D-M+2m$. Since $0\leq m\leq M$, we obtain  the constraint 
\begin{equation}
D-M\leq X\leq D+M. 
\end{equation}
Since the electron moves by two sites when a singlet pair is transmitted  from one side to the other, for $M$ pairs there are $M+1$ allowed values of $X$. This means that, rather than extending over the entire chain,  the electron wave function  is  bounded as illustrated in Fig.  \ref{fig:single-electron-excitation}.

In the following it will be convenient to treat the number of pairs to the left of the electron as a function of the electron position, $m=m(X)=(X-D+M)/2$. 
Using a Schmidt decomposition, we can write a state in the subspace with fixed $D$ and $M$ as\begin{equation}
\ket{\Psi} =\sum_X \ket{X}\otimes \sum_{\nu}\phi_\nu (X)\ket{\psi_{\nu L};X}\otimes \ket{\psi_{\nu R};X},
\end{equation} 
where $\ket{X}$ is the state with a single electron at site $X$,  $\{\ket{\psi_{\nu L};X}\}$ is a  complete orthonormal basis for the Hilbert space with $m(X)$ singlet pairs in the region to  the left of site $X$, and $\{\ket{\psi_{\nu R};X}\}$ is the basis for the Hilbert space with $M-m(X)$ singlet pairs on the right.  The left and right partitions contain $X-1$ and $L-X$ sites, respectively. The coefficients  $\phi_\nu (X)$ can be interpreted as the corresponding single-electron wave functions. 
 

\subsection{Adiabatic approximation}

So far all the manipulations have been exact. To make progress, we note that  the single electron  can be considered as a slow parameter since it only acquires dynamics from the surrounding pairs. Motivated by this observation, we aim for a Born-Oppenheimer-type approximation and factorize the total wave function as
\begin{equation}
\Psi(X;x_{1},\dots,x_{M})=\phi(X)\Phi(x_{1},\dots, x_{M}|X),\label{factorizePhi}
\end{equation}
where $\Phi(x_{1},\dots, x_{M}|X)$  is a  product of two wave functions,
\begin{equation}\label{PhifLfR}
\begin{split}
&\Phi(x_{1},\dots,x_{m},x_{m+1},\dots,x_{M}|X)\\
&\hspace{1cm}=f_{L}(x_{1},\dots,x_{m}|X)f_{R}(x_{m+1},\dots,x_{M}|X).
\end{split}
\end{equation}
Here $f_{L}$ and $f_{R}$ correspond, respectively, to the wave function for singlet pairs to the left and to the right of the single electron. Due to the open  boundary conditions and  the no-double-occupancy constraint, these functions must  satisfy  
\begin{equation}
f_{L}(1/2,\dots,x_{m}|X)=f_{L}(x_{1},\dots,X-1/2|X)=0,
\end{equation}
and
\begin{equation}
f_{R}(X+1/2,\dots,x_{M}|X)=f_{R}(x_{m+1},\dots,L+1/2|X)=0.
\end{equation}

\begin{figure}
\flushleft
\includegraphics[scale=.9]{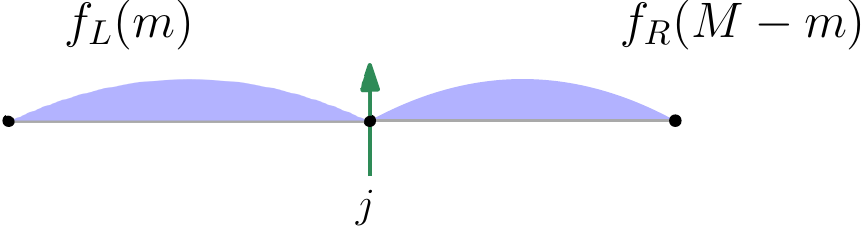}
\includegraphics[scale=.9]{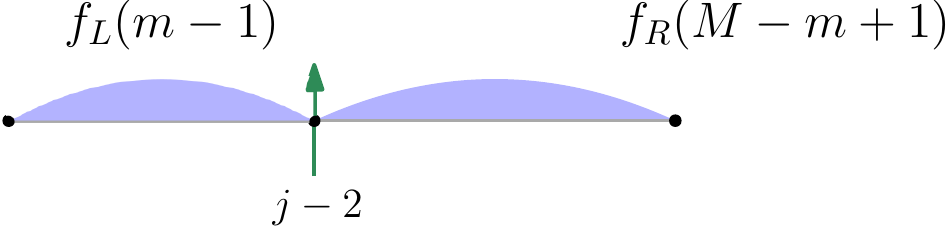}
\caption{Sketch of the adiabatic approximation scheme. At each position of the single electron, the chain is subdivided into two smaller chains. Our ansatz is then constructed by taking the ground state wave function of the singlet-pair liquid in  each box. The  effective length and the number of singlet pairs for each box change  when a singlet pair tunnels across the single electron. }
\label{fig:adiabatic_approx}
\end{figure}

The   integrable $(M,0,0)$ subspace of the theory provides us  with a complete basis of states in which the functions $f_{L}$ and $f_{R}$ can be expanded. We   now   introduce our \emph{de facto} approximation scheme.  Motivated by the adiabatic theorem, we replace  $f_{L}$ and $f_{R}$  by  the lowest-energy states in each Hilbert space. The idea is that the state of the singlet-pair liquid on either side adjusts adiabatically  to the position of the single electron as depicted in Fig. \ref{fig:adiabatic_approx}. In practice, we take
\begin{align}\label{eq:M10-left-wave-function}
&f_{L}(x_{1},\dots,x_{m}|X)\nonumber\\
&\hspace{1.5cm}=\mathcal{N}_{L}\det\left(\sin\big[k_{i}(x_{j}-j+1/2)\big]\right),
\end{align}
where $\mathcal{N}_{L}$ is a normalization factor and $k_{i}$ are quasimomenta given by
\begin{equation}
k_{j}=\frac{\pi}{D-M+m}I_{j},\quad I_{j}=1,\dots,m.
\end{equation}
Likewise, the wave function for the right chain  is
\begin{align}\label{eq:M10-right-wave-function}
&f_{R}(x_{m+1},\dots,x_{M}|X)\nonumber\\
&\hspace{1cm}=\mathcal{N}_{R}\det\left(\sin\big[q_{i}(x_{m+j}-X-j+1/2)\big]\right),
\end{align}
with quasimomenta $q_{i}$ given  by
\begin{equation}
q_{j}=\frac{\pi}{L-D-m+1}I_{j},\quad I_{j}=1,\dots,M-m.
\end{equation}
In particular, the adiabatic approximation becomes exact for the ground state in the $(M,1,0)$ subspace, since in this case the electron sits at one of the boundaries  (such that $D=M+1$ or $D=L-M$) and decouples from the dynamics of the singlet pairs moving in the remaining sites.


\subsection{Effective Hamiltonian for the single electron}

Our goal now is to determine the single-electron  wave function $\phi(X)$ within the adiabatic approximation. For this purpose, we substitute our ansatz given by Eqs. (\ref{factorizePhi}), (\ref{PhifLfR}), (\ref{eq:M10-left-wave-function}) and (\ref{eq:M10-right-wave-function}) into the Schr\"odinger equation and sum out the singlet-pair degrees of freedom.  The analog of Eq. (\ref{eq:110-meet}) in this case is  an inhomogeneous tight-binding Hamiltonian:
\begin{align}\label{eq:M10-eff-single-hop}
&\big[E-u(X)\big]\phi(X)\nonumber\\
&\hspace{1.5cm}=t_{-}(X)\phi(X-2)+t_{+}(X)\phi(X+2),
\end{align}
with boundary conditions $\phi(D+M+2)=\phi(D+M-2)=0$. The effective hopping parameters $t_+$ and $t_-$  depend on the overlap between singlet-pair wave functions for different  electron positions:
\begin{eqnarray}
\hspace{-.4cm}t_{\pm}(X)&=&\gamma \sum_{\{x_{i}\}}{}'\Phi^{*}(x_{1},\dots,x_{m}=X\pm3/2,\dots,x_{M}|X) \nonumber\\
&&\times\Phi(x_{1},\dots,x_{m}=X\mp1/2,\dots,x_{M}|X\pm2).
\end{eqnarray}
Here $\gamma=1/2$ and the primed sum is performed over all allowed values of $x_j$, with exception of $x_{m}$ which  is kept fixed. The on-site potential term $u(X)$ appearing in Eq. (\ref{eq:M10-eff-single-hop}) is the sum of the ground state energies for the disconnected chains:
\begin{equation}
u(X)=E_{g}(m;X-1)+E_{g}(M-m;L-X),
\end{equation}
where $E_{g}(M;L)$ is the ground state energy for $M$ pairs on a chain with $L$ sites,
\begin{align}
&E_{g}(M;L)\nonumber\\
&\hspace{.2cm}=1-\csc\left(\frac{\pi}{2L-2M+2}\right)\sin\left[\frac{\pi(2M+1)}{2L-2M+2}\right].
\end{align}

\begin{figure}
\includegraphics[width=\columnwidth]{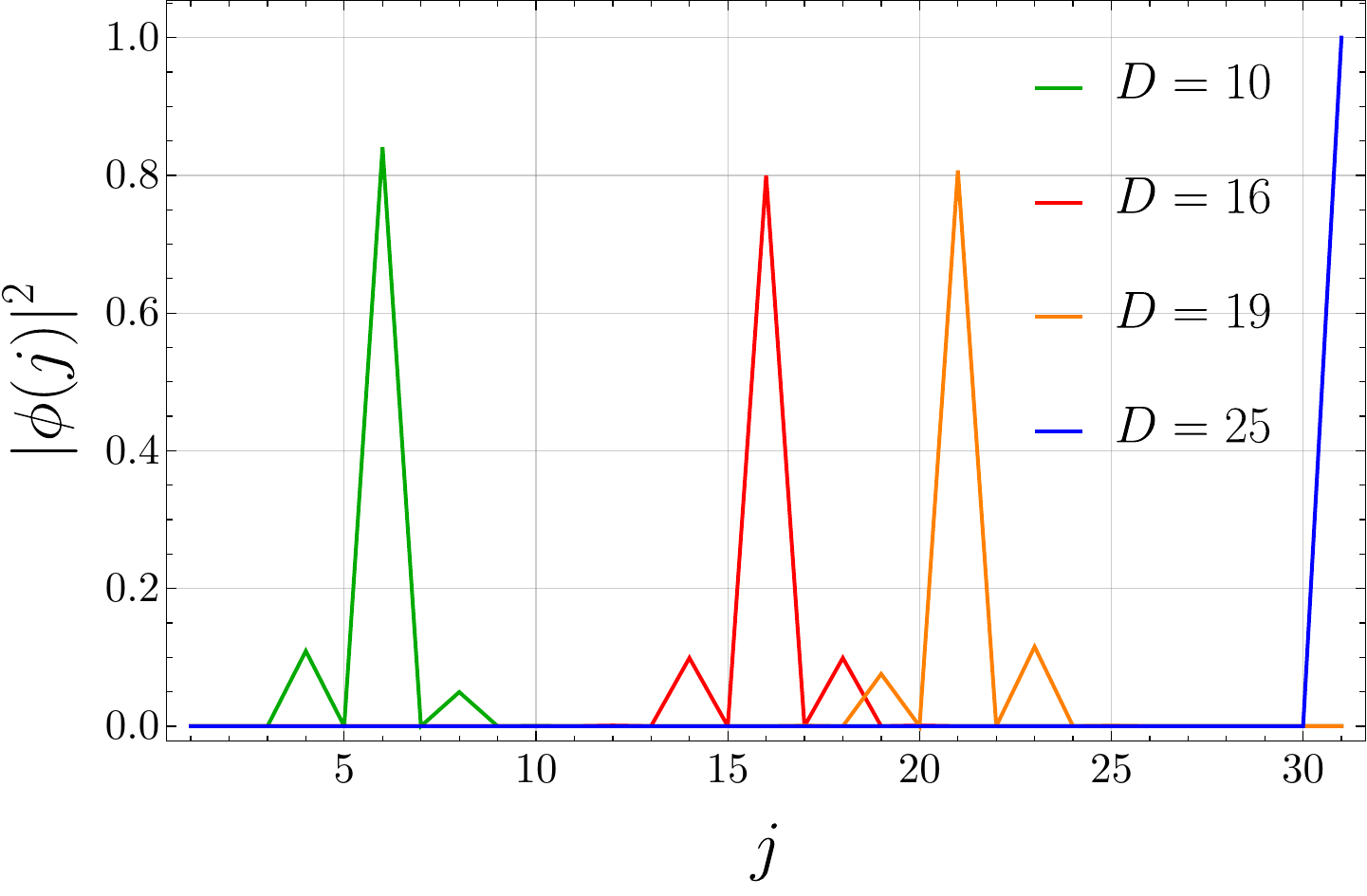}
\caption{Single-electron probability density calculated  in the adiabatic approximation for a system with $M=6$ singlet pairs in a chain with $L=31$ sites. The different curves correspond to different values of the  dipole moment $D$.}
\label{fig:ElectronWaveFunction}
\end{figure}

We then solve the system of Eqs. (\ref{eq:M10-eff-single-hop}) numerically for a fixed choice of parameters $M$, $D$ and $L$. In Fig. \ref{fig:ElectronWaveFunction}, we plot the  probability density $|\phi(j)|^2$ corresponding to the ground state of the effective Hamiltonian for a single electron interacting with $M=6$ singlet pairs in a chain with $L=31$ sites.
We can see that the single electron becomes restricted to a few sites and the probability density is skewed to the nearest boundary. 
This shows that the same effects observed before for just one singlet pair (see Figs. \ref{fig:GS_energy_110_wire} and \ref{fig:ElectronWaveFunction-110}) remain true for a larger number of pairs.

We stress that the disorder-free localization of single electrons encountered here is a direct consequence of the dipole conservation  law. The latter introduces constraints, such as the one   illustrated in Fig. \ref{fig:single-electron-excitation}, that decouple distinct subspaces and lead to an extensive fragmentation of the Hilbert space.
This mechanism --- known as Krylov fragmentation \cite{Hudomal2020,Moudgalya2020} --- has been identified in other dipole conserving models \cite{Sala2020} and can give rise to  quantum scarred subspaces  that violate the eigenstate thermalization hypothesis \cite{Serbyn2020,Khemani2020}. 
Therefore, the singlet pair liquid with a finite density of single electrons   provides another setting for   studies of ergodicity breaking in translation-invariant systems \cite{Smith2017,Smith2019,Hart2020}.


\section{Perturbations}\label{sec:perturbations}

In this section we ask what sort of couplings can be switched on to close energy   gaps and induce quantum phase transitions in the singlet-pair hopping model. Here we have chosen to study the effects of a magnetic field and a nearest-neighbor repulsion. Our particular choice has been made with two conditions in mind. First, both interactions commute with the dipole operator in Eq. (\ref{eq:dipole-operator}), preserving fracton physics. Second, they do not couple spaces with distinct occupation numbers in the classification of Sec. \ref{sec:model-charges}, such that we can solve everything in the same way as before. Single-particle hopping, which does not fulfill any of our conditions, is briefly discussed in Appendix \ref{app:hopping}.


\subsection{Magnetic field}\label{perturbh}

Given that   Hamiltonian (\ref{eq:model})   commutes with $S^z$, coupling the system to an external magnetic field $h$ in the $z$-direction is   the simplest way to close the spin gap. Let us then consider the modified Hamiltonian
\begin{equation}\label{eq:}
H =\mathcal{P}\Big[-\sum_{j}\left(b_{j-1,j}^{\dagger}b_{j,j+1}+\text{H.c.}\right)-h\sum_{j} S^{z}_{j}\Big]\mathcal{P},
\end{equation} 
where we assume $h>0$. All eigenstates   remain the same as before. The only effect is a shift on the eigenenergies according to
\begin{equation}
E(N,S^z)\rightarrow E(N,S^z)-hS^z,
\end{equation}
where $E(N,S^z)$ is an eigenenergy of   Hamiltonian (\ref{eq:model}) with $N$ electrons and total spin $z$-projection $S^z$.
As a consequence, the spin gap closes when there is a level crossing at the critical magnetic field  $h_c$ given by [see Eq. (\ref{eq:spin-gap})]
\begin{equation}
h_c= 1 + \cos\left(\frac{\pi n}{2-n}\right).
\end{equation}

\begin{figure}
\includegraphics[width=\columnwidth]{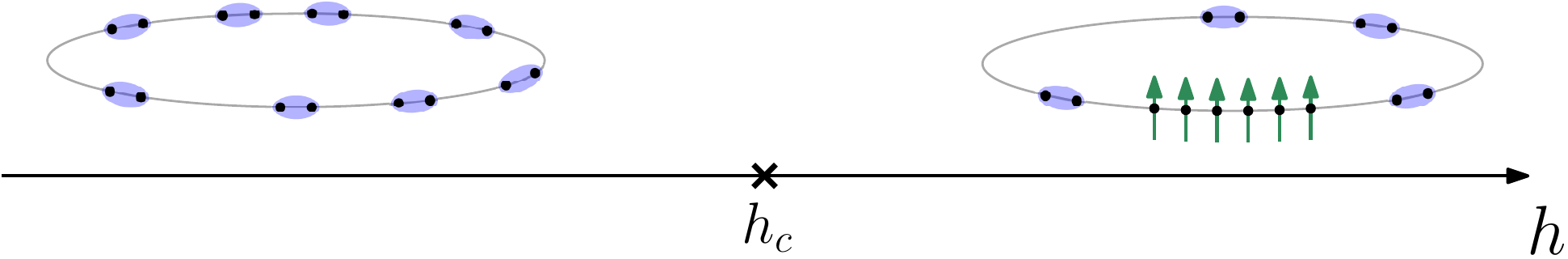}
\caption{Phase diagram for the model with   an external  magnetic field. Above the critical point, $h>h_c$, an island of polarized electrons emerges in the ground state.}
\label{fig:cluster_formation}
\end{figure}

For $h>h_c$, we expect  phase separation with the formation  of ``electronic islands" as illustrated in Fig. \ref{fig:cluster_formation}. The mechanism behind these polarized electronic clusters can be understood as follows. Slightly above the critical field, the ground state with a fixed number of electrons  in a finite chain    changes from $S^z=0$ to $S^z=1$ by turning a singlet pair into a  triplet pair. The latter then  acts as a hard wall for the remaining singlet pairs. If we increase the field further, any new excitations become attracted to the triplet pair as suggested by Figs. \ref{fig:GS_energy_110_wire}, \ref{fig:ElectronWaveFunction-110}, \ref{fig:GS-energy-120}, and \ref {fig:ElectronWaveFunction}. To make this idea more concrete, we can compare the energies of two states in the $S^z=2$ sector. In the first state, we put the   triplet pairs   together,  forming a four-electron cluster. In the second state, we place the two triplet pairs at a maximum distance from each other inside the finite-length ring.  This problem  is equivalent to considering one and two interval partitions of the ring, respectively. Calculating the energy difference between the two states, we determine the energy necessary to separate the   triplet pairs: 
\begin{eqnarray}
\Delta E &=& 1 +\frac{2}{\pi}\sin\left(\frac{\pi n}{2-n}\right)\nonumber\\
&&-\left(\frac{2-3n}{2-n}\right)\cos\left(\frac{\pi n}{2-n}\right)+O(1/L).
\end{eqnarray}
Thus, there is a finite energy gap for breaking the cluster. This supports the reasoning that for $h>h_c$ the ground state on the ring contains a cluster of polarized electrons surrounded by a liquid of singlet pairs.


\subsection{Nearest-neighbor repulsion}

One route to closing the one-particle excitation gap  is to penalize singlet pairs by introducing  a nearest-neighbor repulsive interaction. In this scenario, the  Hamiltonian takes the form
\begin{equation}\label{eq:model-V}
H =\mathcal{P}\Big[-\sum_{j}\left(b_{j-1,j}^{\dagger}b_{j,j+1}+\text{H.c.}\right)+V\sum_{j}n_{j}n_{j+1}\Big]\mathcal{P},
\end{equation}
with $V> 0$. Unlike   the magnetic field considered in Sec. \ref{perturbh},  the nearest-neighbor interaction  does not commute with   pair hopping   and we need to find new solutions to Eq. (\ref{eq:model-V}). Fortunately, the new term   preserves the integrability of the $(M,0,0)$ subspaces, and the resulting Bethe equations are analogous to those for the  anisotropic (XXZ) spin-$1/2$ chain  \cite{Sutherland2004,Zvyagin2005}.

In the $(M,0,0)$ subspace, we try the following wave function for $M$ singlet  pairs:
\begin{equation}\label{eq:M00-BA-wave-function-V}
\Psi(x_{1},\dots,x_{M})=\sum_{P}A(P)\exp\Bigg(i\sum_{j=1}^{M}k_{Pj}x_{j}\Bigg),
\end{equation}
where the summation $P$ runs over all permutations and negations of the momenta $(k_{1},\dots,k_{M})$. The energy of such a state is given by
\begin{equation}\label{eq:energy-momentum-M00-V}
E[k]=MV-2\sum_{j=1}^{M}\cos k_{j}.
\end{equation}
The singlet pair scattering matrix is   modified to
\begin{equation}\label{eq:singlet-pair-S-matrix-V}
S(k,k')=-e^{-i(k-k')+i\vartheta(k,k')},
\end{equation}
where $\vartheta(k,k')$ is the two-body scattering phase shift due to the interaction potential $V$,
\begin{equation}
\vartheta(k,k')=-i\log\left[\frac{1+e^{i(k+k')}+Ve^{ik'}}{1+e^{i(k+k')}+Ve^{ik}}\right].
\end{equation}
The quantization condition for an open chain with $L$ sites takes the   form
\begin{eqnarray}
1 & =&S(k_{j},k_{j+1})\dots S(k_{j},k_{M})s_{L+1/2}(k_j,-k_j)S(k_{M},-k_{j})\nonumber\\
&&\times \dots  S(k_{j+1},-k_{j})S(k_{j-1},-k_{j})\dots S(k_{1},-k_{j})\nonumber\\
 & &\times s_{1/2}(-k_j,k_j)S(k_{1},k_{j})\dots S(k_{j-1},k_{j}),
\end{eqnarray}
where $s_{x}(k,-k)=-e^{2ikx}$ denotes the scattering amplitude at reflective boundaries as defined in Eqs. (\ref{eq:scatter-reflection-left}) and (\ref{eq:scatter-reflection-right}).
Using   $S(k,q)S(q,-k)=e^{ -2ik+i\left[\vartheta(k,q)-\vartheta(q,-k)\right]}$, we obtain   the Bethe equations  
\begin{equation}\label{eq:Bethe-eq-M00-V}
2k_{j}(L-M+1)+\sum_{j\neq l}\big[\vartheta(k_{j},k_{l})-\vartheta(k_{l},-k_{j})\big]=2\pi I_{j},
\end{equation}
where $I_j$ is an integer that parametrizes the branch of the  logarithm. Within   the $(M,0,0)$ sector  the  nearest-neighbor repulsion   renormalizes the Luttinger parameter for the singlet-pair liquid, in analogy with the effect of the exchange anisotropy in the XXZ spin chain \cite{Sutherland2004}.

We now turn to the solution in the $(M,1,0)$ subspace. As discussed in Sec. \ref{sec:singleparticle}, to find the ground state in this sector we  just  place  a single electron at one of the boundaries, say $j=L$. The presence of the single electron  changes the effective length of the chain, and  the scattering amplitude becomes
\begin{equation}
s_{L-1/2}(k,-k)=-e^{ik(2L-1) +i \varphi(k)},
\end{equation}
where $\varphi(k)$ is the contribution  from the interaction potential  to the scattering phase shift between a singlet pair  and the electron,
\begin{equation}
\varphi(k)=-i\log\left(\frac{1+Ve^{-ik}}{1+Ve^{ik}}\right).
\end{equation}
Hence, we find   the Bethe equations in this subspace:  
\begin{align}\label{eq:Bethe-eq-M10-V}
&2k_{j}(L-M)+\varphi(k_{j})\nonumber\\
&\hspace{1cm}+\sum_{j\neq l}\big[\vartheta(k_{j},k_{l})-\vartheta(k_{l},-k_{j})\big]=2\pi I_{j}.
\end{align}

\begin{figure}
\includegraphics[width=\columnwidth]{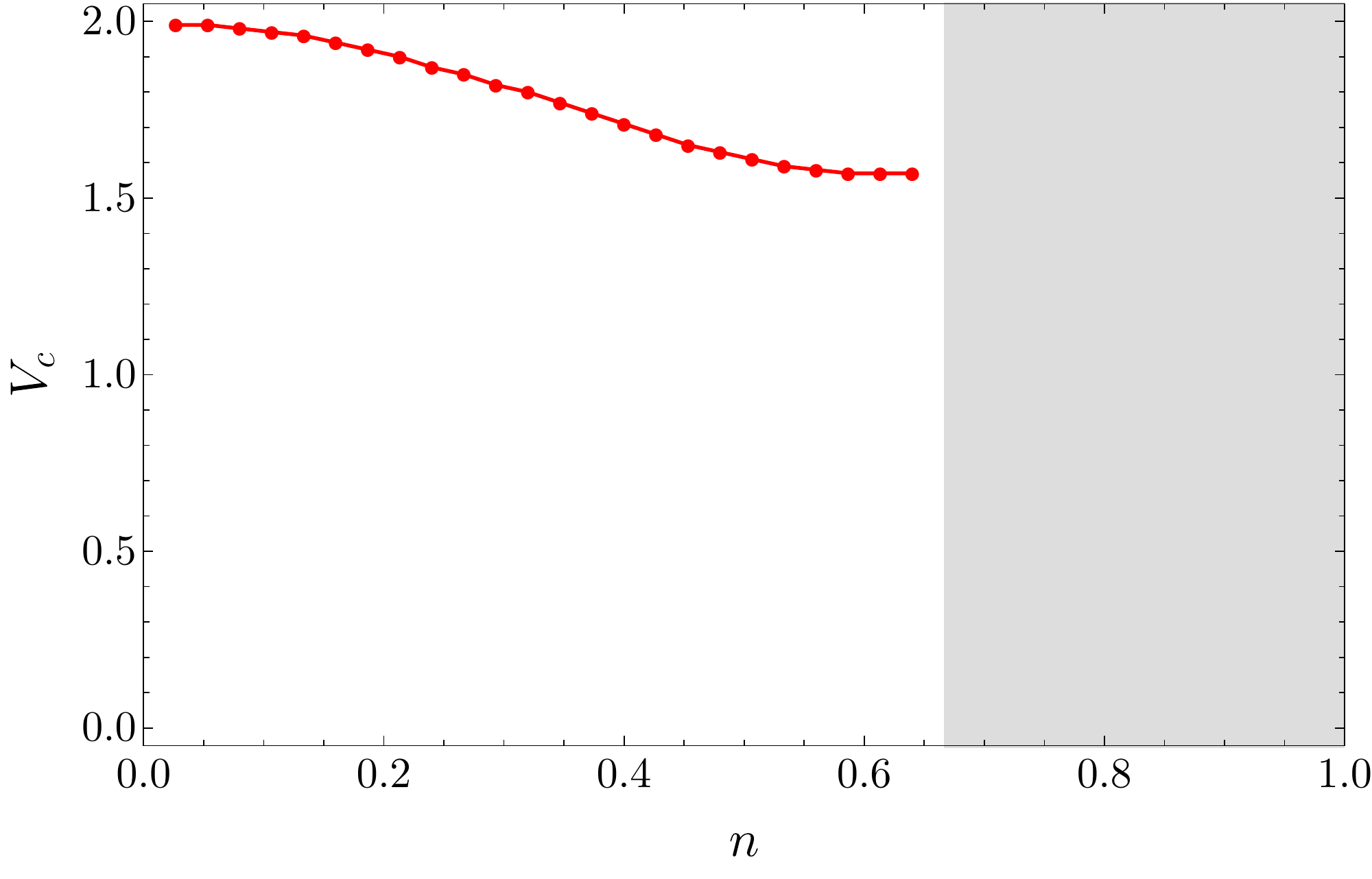}
\caption{Critical repulsion $V_c$ as a function of electronic filling $n$. For $n>2/3$, the nearest-neighbor interaction does not modify the one-particle excitation gap. }
\label{fig:Vc}
\end{figure}

We have  solved  Eqs. (\ref{eq:Bethe-eq-M00-V}) and (\ref{eq:Bethe-eq-M10-V}) numerically for a chain with $L=75$ sites.  This way we are able to  determine the critical interaction strength  $V_c$ required to close the one-particle gap $\Delta_{1p}$. The result is shown in Fig. \ref{fig:Vc}. For electronic fillings   $n<2/3$, we find a finite value of $V_c$ separating the singlet-pair liquid phase, with gapped single-electron excitations, from a phase with a finite density of fractons. Lacking exact solutions, we cannot ascertain the nature of the phase  for $V>V_c$. However, we speculate that the interplay between nearest-neighbor repulsion and long-range attraction between electrons may stabilize a crystalline phase with periodic arrangement of fractons within a cluster. Finally, for $n>2/3$ we find that the nearest-neighbor interaction does not modify the one-particle gap. In fact, at the commensurate filling $n=2/3$, due to relevant Umklapp scattering interactions we antecipate that an infinitesimal $V>0$ turns the singlet pair liquid into an insulating valence bond crystal. The situation is similar to the charge density wave instability of hardcore spinless fermions \cite{Gomez-Santos1993}.


\section{Conclusions}\label{sec:conclusions}

We investigated a liquid of singlet pairs in which single-electron excitations behave as fractons. Our results were established in a one-dimensional model of interacting electrons that describes hopping of spin-singlet pairs. We showed that the model obeys a dipole moment conservation law and explored its Bethe ansatz integrable spaces to obtain several exact results. We introduced an adiabatic approximation that allowed us to capture the essential physics of the localized wave function of the single electron immersed in the pair liquid. We also studied two perturbations that are able to close the gaps for spin and single-particle excitations, observing the important role played by the effective attraction between fractons in the resulting phases.

Future directions  include   investigating  the out-of-equilibrium properties of the system. One advantage of our model in comparison with spinless fermion models is that single electrons immersed in the singlet pair liquid  could  be  detected by measuring  the expectation value of the local spin projection. Thus, one could probe the magnetization dynamics which should be characterized by the absence of spin diffusion. 
Moreover, it may be possible to extend the low-energy field theory approach to study the single-electron dynamics as a  kinetically constrained quantum impurity problem directly in the thermodynamic limit. This would allow us to investigate the disorder-free localization beyond the adiabatic approximation.  It would also be  worth investigating the relation between the emergence of correlated hopping and the strong-coupling limit of gauge theories coupled to matter fields \cite{Borla2020}.
In particular, it has been shown that a one-dimensional model of spinless fermions that includes pair-hopping terms can host topological edge modes \cite{Ruhman2017}, which might lead to a connection between topological and fractonic behavior even in one spatial dimension.
Given the simple microscopic mechanism behind the singlet-pair hopping Hamiltonian, mixed-dimensional Mott insulators \cite{Grudst2018} and Floquet enginnering in optical lattices \cite{Coulthard2018,Gao2020} may open the possibility to simulate such models in the future.
Furthermore, since the three-site correlated hopping promotes an electronic pairing similar to the one expected from Cooper pairs, one may wonder about its role in the onset of superconductivity in higher dimensions.


\begin{acknowledgments}

We acknowledge  financial support from CAPES (HBX) and  CNPq (RGP). Research at IIP-UFRN is funded by the Brazilian ministries MEC and MCTI.

\end{acknowledgments}


\appendix


\section{Single particle hopping}\label{app:hopping}

Single-particle hopping manifestly breaks the dipole moment Eq. (\ref{eq:dipole-operator}) conservation law, indicating a departure from fracton physics. It is unclear whether some sort of approximate fracton behavior can still persist for perturbatively small single particle-hopping amplitudes. However, the existence of singlet-pair bound states separated from a continuum of  scattering states can be demonstrated by directly solving the two-electron problem on the lattice.

If we switch on single-particle hopping, Hamiltonian (\ref{eq:model}) will become
\begin{align}\label{eq:t-aJ-model}
H &=\mathcal{P}\Big[-\sum_{j}\left(b_{j-1,j}^{\dagger}b_{j,j+1}+\text{H.c.}\right)\nonumber\\
&\hspace{1cm}-t\sum_{j\sigma}\left(c_{j,\sigma}^{\dagger}c_{j+1,\sigma}+\text{H.c.}\right)\Big]\mathcal{P},
\end{align}
where we assume $t>0$. Two-electron eigenstates take the form 
\begin{equation}
\ket{\psi} =
\sum_{\sigma_{1},\sigma_{2}}\sum_{x_{1},x_{2}}\psi(x_{1}\sigma_{1},x_{2}\sigma_{2})c_{x_{1}\sigma_{1}}^{\dagger}c_{x_{2}\sigma_{2}}^{\dagger}\ket{0} ,
\end{equation}
where the wave function is totally antisymmetric $\psi(x_{1}\sigma_{1},x_{2}\sigma_{2})=-\psi(x_{2}\sigma_{2},x_{1}\sigma_{1})$.

\begin{figure}
\includegraphics[width=\columnwidth]{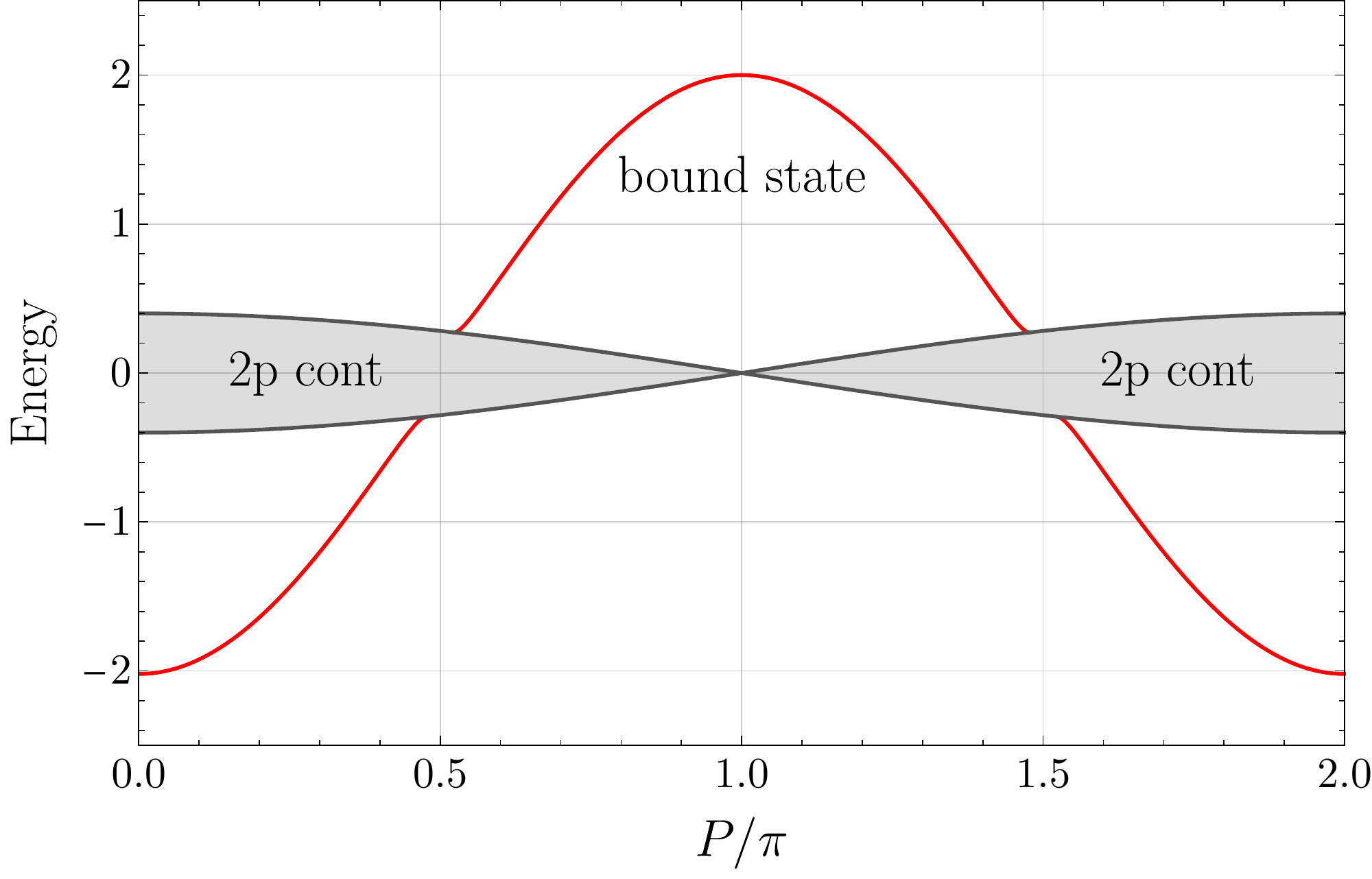}
\caption{Two-particle continuum and bound state bands in the spectrum of two-body problem including single-particle hopping $t$ as a perturbation. Here we set  $t=0.1$ in   units of the pair hopping amplitude.}
\label{fig:two-body-bound-state}
\end{figure}

We now look for solutions of the Schr\"odinger equation for the two-body problem. When electrons are well separated, $|x_{1}-x_{2}|>1$, we have 
\begin{align}\label{eq:t-hop-2e-eq-separated}
E\psi(x_{1},x_{2})&=-t\psi(x_{1}+1,x_{2})-t\psi(x_{1}-1,x_{2})\nonumber\\
&\quad-t\psi(x_{1},x_{2}+1)-t\psi(x_{1},x_{2}-1).
\end{align}
On the other hand, when the electrons meet, there is also the possibility to pair them up
\begin{align}\label{eq:t-hop-2e-eq-meet}
E\psi&(x,x+1)=-t\psi(x-1,x+1)-t\psi(x,x+2) \nonumber\\
& -\frac{1}{2}\left(1-\Pi\right)\left[ \psi(x,x-1)+\psi(x+2,x+1)\right],\\
E\psi&(x,x-1)=-t\psi(x+1,x-1)-t\psi(x,x-2) \nonumber\\
& -\frac{1}{2}\left(1-\Pi\right)\left[ \psi(x-2,x-1)+\psi(x,x+1)\right] .\nonumber
\end{align}
Here we omit spin variables for brevity and have introduced the operator $\Pi$ that interchanges spin variables, i.e.,
\begin{equation}
\Pi\psi(x_{1}\sigma_{1},x_{2}\sigma_{2})=\psi(x_{1}\sigma_{2},x_{2}\sigma_{1}).
\end{equation}
In the most general form, the spin wave function of $\psi(x_{1},x_{2};\sigma_{1},\sigma_{2})$ is written as a linear combination of the spin singlet state
\begin{equation}
\varphi_{0}(\sigma_{1},\sigma_{2})=\delta_{\sigma_{1},\uparrow}\delta_{\sigma_{2},\downarrow}-\delta_{\sigma_{1},\downarrow}\delta_{\sigma_{2},\uparrow}
\end{equation}
and the spin triplet states
\begin{equation}
\varphi_{1}(\sigma_{1},\sigma_{2})=
\begin{cases}
\delta_{\sigma_{1},\uparrow}\delta_{\sigma_{2},\uparrow}, & S^{z}=1,\\
\delta_{\sigma_{1},\uparrow}\delta_{\sigma_{2},\downarrow}+\delta_{\sigma_{1},\downarrow}\delta_{\sigma_{2},\uparrow}, & S^{z}=0,\\
\delta_{\sigma_{1},\downarrow}\delta_{\sigma_{2},\downarrow}, & S^{z}=-1.
\end{cases}
\end{equation}
The spin singlet state is antisymmetric in the spin variables and symmetric in the electron coordinates, while the situation is reversed for the spin triplet states. Moreover, the spin triplet states do not couple to the singlet-pair hopping term and the problem becomes identical to  free spinless fermions. In view of that, we   search for bound states restricting the analysis to the spin singlet state. As usual, we may separate center of mass and relative coordinates as
\begin{equation}\label{eq:t-hop-2e-ansatz}
\psi(x_{1}\sigma_1,x_{2}\sigma_2)=\varphi_0(\sigma_1,\sigma_2) e^{iP(x_{1}+x_{2})/2}f(x_{1}-x_{2}),
\end{equation}
where $P$ is the total momentum and $f(r)$ is an even function so that $f(-r)=f(r)$. Plugging   Eq. (\ref{eq:t-hop-2e-ansatz}) into the set of Eqs. (\ref{eq:t-hop-2e-eq-separated}) and (\ref{eq:t-hop-2e-eq-meet}) yields
\begin{equation}\label{eq:t-hop-2e-eq-separated-r}
Ef(r)=-2t\cos(P/2)\left[f(r+1)+f(r-1)\right],
\end{equation}
when the electrons are separated, and
\begin{equation}\label{eq:t-hop-2e-eq-meet-r}
Ef(1)=-2t\cos(P/2)f(2)-2\cos(P)f(-1)
\end{equation}
when they meet. Bound states correspond to solutions of the form $f(r)=e^{-\kappa |r|}$ with  real $\kappa>0$. Equation (\ref{eq:t-hop-2e-eq-separated-r}) determines the energy,
\begin{equation}
E=-4t\cos(P/2)\cosh(\kappa).
\end{equation}
On the other hand, Eq. (\ref{eq:t-hop-2e-eq-meet-r}) determines  $\kappa$ as a function of the total momentum:
\begin{equation}
e^{-\kappa}=t\left(\frac{\cos P/2}{\cos P}\right).
\end{equation}
Thus, given that both $\kappa$ and $t$ are real and positive, we   find two possibilities:  ({\it i}) if $t<1$, there is  a bound state with momentum  near $P=0$ and an anti-bound state near $P=\pi$; ({\it ii})   if $t>1$,  there is only an anti-bound state centered at $P=\pi$. In Fig. \ref{fig:two-body-bound-state} we plot the solution for case ({\it i}). The dispersion relation for this two-particle bound state is given by
\begin{equation}
E=-2\Bigg[ 1+t^{2}\left(\frac{\cos P/2}{\cos P}\right)^{2}\Bigg] \cos P.
\end{equation}
As a check, note  that in the limit $t\rightarrow0$, for $P\neq\pm \pi/2$, we recover $E=-2\cos P$, which is the solution for one singlet pair with momentum $P$. This solution of the two-electron problem suggests that the binding of electrons into singlet pairs may persist in the ground state of the model with small $t$. However, the conservation of the dipole moment is immediately broken for any $t\neq 0$. In this case, we expect the single-electron wave functions  to become extended.


\bibliography{references}


\end{document}